\documentclass[acmsmall]{acmart}

\AtBeginDocument{%
  }

\setcopyright{acmcopyright}
\copyrightyear{2023}
\acmYear{2023}
\acmDOI{XXXXXXX.XXXXXXX}

\acmJournal{TOIS}
\acmVolume{37}
\acmNumber{4}
\acmArticle{111}
\acmMonth{8}

\usepackage{xspace}
\usepackage{algorithmic}
\usepackage{algorithm2e}
\usepackage{multirow}
\usepackage{amsmath}
\usepackage{graphicx}
\allowdisplaybreaks[4]

\newcommand{\eat}[1]{}
\newcommand{\paratitle}[1]{\smallskip\noindent \textbf{#1}}
\newcommand{\ie}{\emph{i.e.,}\xspace}
\newcommand{\eg}{\emph{e.g.,}\xspace}
\newcommand{\etal}{\emph{et al.}\xspace}

\newcommand{\baby}{\textsc{DiffuRec}\xspace}

\begin{document}
\title{DiffuRec: A Diffusion Model for Sequential Recommendation}

\author{Zihao Li}
\email{zihao.li@whu.edu.cn}
\orcid{0000-0002-1802-4284}
\author{Chenliang Li}
\orcid{0000-0003-3144-6374}
\authornote{Chenliang Li is the corresponding author.}
\email{cllee@whu.edu.cn}
\affiliation{%
  \institution{Key Laboratory of Aerospace Information Security and Trusted Computing, Ministry of Education, School of Cyber Science and Engineering, Wuhan University}
  \streetaddress{299 Bayi Road, Wuchang District}
  \city{Wuhan}
  \state{Hubei}
  \country{China}
  \postcode{430072}
}

\author{Aixin Sun}
\orcid{0000-0003-0764-4258}
\affiliation{%
  \institution{Nanyang Technological University}
  \country{Singapore}
  }
\email{axsun@ntu.edu.sg}


\renewcommand{\shortauthors}{Z. Li et al.}

\begin{abstract}
  Mainstream solutions to Sequential Recommendation (SR) represent items with fixed vectors. These vectors have limited capability in capturing items' latent aspects and users' diverse preferences. As a new generative paradigm, \textit{Diffusion models} have achieved excellent performance in areas like computer vision and natural language processing. To our understanding, its unique merit in representation generation well fits the problem setting of sequential recommendation. In this paper, we make the very first attempt to adapt Diffusion model to SR and propose \baby, for item representation construction and uncertainty injection. Rather than modeling item representations as fixed vectors, we represent them as distributions in \baby, which reflect user's multiple interests and item's various aspects adaptively. In diffusion phase, \baby corrupts the target item embedding into a Gaussian distribution via noise adding, which is further applied for sequential item distribution representation generation and uncertainty injection. Afterward, the item representation is fed into an Approximator for target item representation reconstruction. In reverse phase, based on user's historical interaction behaviors, we reverse a Gaussian noise into the target item representation, then apply a rounding operation for target item prediction. Experiments over four datasets show that  \baby outperforms strong baselines by a large margin\footnote{Code is available at: https://github.com/WHUIR/DiffuRec}.
\end{abstract}



\received{20 February 2007}
\received[revised]{12 March 2009}
\received[accepted]{5 June 2009}

\maketitle

\begin{figure}[t]
\centerline{\includegraphics[width=0.7\textwidth]{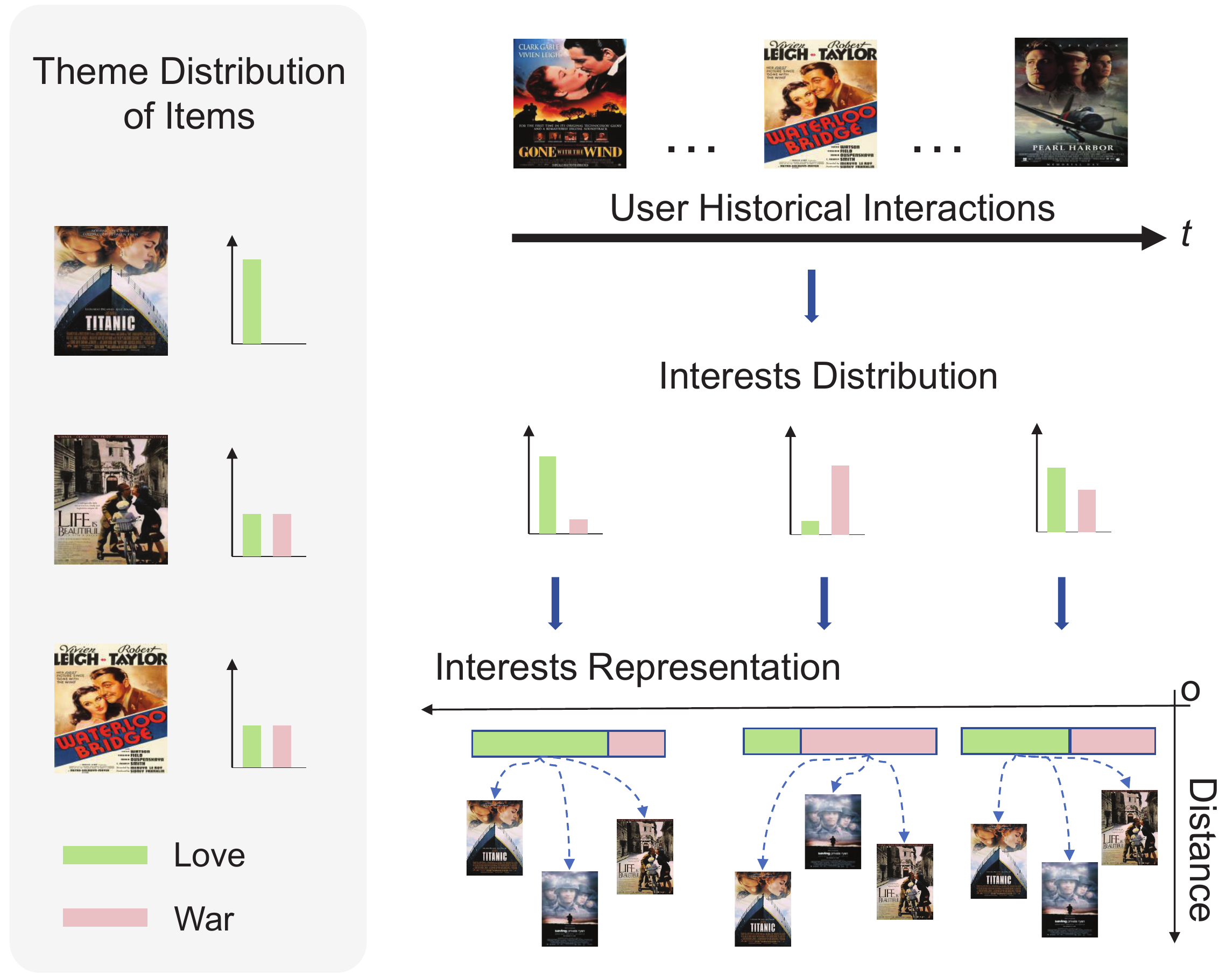}}
\caption{An example of multiple interests of users and multiple aspects of items.}
\label{fig:toy_example}
\end{figure}

\section{Introduction}
\label{sec:intro}

Sequential Recommendation (SR) aims to model a user's interest evolution process, based on his/her historical interaction records, for the next item prediction. The task has attracted widespread attention recently due to its significant business value. 
Seen as a sequence modeling task, sequential neural networks like LSTM have been widely adopted in early studies~\cite{hidasi2018recurrent, hidasi2015session}. Recent efforts illustrated the remarkable ability of self-attention mechanism for long-term dependency capturing.  Thus, self-attention based methods became a prominent solution~\cite{sun2019bert4rec, kang2018self, wu2020sse, yuan2020future}. SASRec~\cite{kang2018self} is a pioneering work that applies a uni-direction Transformer to learn transition patterns of items in sequences. BERT4Rec~\cite{sun2019bert4rec} adopted a bi-direction Transformer encoder to predict the next item. Based on information propagation and aggregation, Graph Neural Network (GNN) has also achieved impressive results on SR~\cite{guo2021gcn, fan2021continuous}. Compared to earlier solutions, GNN could capture high-order dependencies between items.


All these mainstream methods learn item representation as an embedding vector. However, we \textcolor{black}{believe} a fixed vector may have limited capability in capturing the following four characteristics simultaneously.

\begin{itemize}
    \item \textbf{Multiple Latent Aspects.} In practical scenarios, items may contain different latent aspects. We take the movie recommendation scenario as a running example (as shown in Figure~\ref{fig:toy_example}). The aspects of items can be regarded as the themes or categories of movies. Thus, different movies will contain different theme distributions (circled on the left). For example, \textit{Titanic} is a romantic movie, while \textit{Life is Beautiful} contains both themes of Love and War. Note that in many recommendation scenarios, it is difficult to clearly define and precisely annotate items with their corresponding aspects. 
    Encoding the complex latent aspects in a \textcolor{black}{single} vector remains challenging\textcolor{black}{~\cite{fan2021modeling, fan2022sequential}}.
    \item \textbf{Multiple Interests.} User's interests and preferences are diverse and inconstant. Further, user may dynamically adjust her interests based on different scenarios or periods. Hence, her focus on the various latent aspects of an item may also drift. As shown in Figure~\ref{fig:toy_example}, when this user's preference to the themes (\ie Love and War) is changed. The importance of each latent aspect for an item should also be adjusted accordingly, in a user-aware manner. 
    From this perspective, a static item representation across users is insufficient as well~\cite{fan2022sequential}.
    \item \textbf{Uncertainty.} Due to the diversity and evolution of user interests, user's current preference and intention become uncertain to some degree. On the other hand, the diversity, novelty, serendipity, and uncertainty reflected through the recommended items are also expected from a recommender system~\cite{zhang2012auralist, fan2021modeling, hurley2011novelty}. We argue that it is more appropriate to model user's current interests as a distribution.
    \item \textbf{Guidance of Target Item.} At last, the target item could facilitate the user's current intention understanding, and it can be introduced as an auxiliary signal to guide the latter process~\cite{wang2022target}. However, since the interaction involving the target item could introduce tremendous computation cost and deep coupling for the network design, the inference speed is also highly related to the size of the candidate pool. Although few works have been proposed, they are only applicable to the ranking stage.
\end{itemize}

To partially address the aforementioned characteristics, attention mechanism and dynamic routing for multi-interest modeling and recommendation have been explored~\cite{zhou2018deep, li2019multi, cen2020controllable, wang2022target}. However, those methods require a proper number of interests (or capsules) to be heuristically pre-defined for model training. Besides, this pre-defined and fixed number also constrains the representation ability and flexibility of models. In addition to these solutions, Variational Auto-Encoder (VAE) methods have the potential to model the multiple latent aspects of an item as a distribution and inject uncertainty to the models via Gaussian noises. However, VAE suffers from the limited representation capacity~\cite{xie2021adversarial} and collapse issue of posterior distribution approximation~\cite{liang2018variational, zhao2019infovae}, precluding the models to achieve the best performance. 
Furthermore, none of existing methods could model the multiple interests of users and multiple latent aspects of items as distributions, under a unified framework. 

Diffusion models have made remarkable success in CV, NLP and many other fields~\cite{avrahami2022blended, li2022diffusion, yang2022diffusion, ramesh2022hierarchical}. With the merits of its distribution generation and diversity representation, we consider Diffusion model to be a good fit to sequential recommendation. In this paper, we thereby make the very first attempt to 
bridge the gap between diffusion model and sequential recommendation, and propose \baby. 

In the proposed \baby, in diffusion phase, we gradually add Gaussian noise into the target item embedding. The noised target item representation is then fed as input to generate distribution representations for the historical items, \ie to exploit the guidance of target item.
\textcolor{black}{As such, the item's multi-latent aspects and user's multi-interests could be represented as distributions instead of fixed vectors and the supervised signal encapsulated in the target item could also be fused for representation generation simultaneously.}
Note that the noised target item representation also injects some uncertainty into this process, leading to more discriminative item embedding learning. 
Then, based on the distribution representations of these historical items, we utilize a Transformer model as \textit{Approximator}, to reconstruct target item representation for model training.  As to the inference, we iteratively reverse a pure Gaussian noise into the target item distribution representation. The well-trained approximator will reveal the user's current interest iteratively. Furthermore, considering the reversed target item representation distributes in a high-dimensional continuous space~\cite{li2022diffusion}, we devise a rounding function to map it into a discrete item embedding space for target item prediction. In a nutshell, we make the following contributions in this paper: 

\begin{itemize}
    \item 
    To the best of our knowledge, this is the first attempt to apply diffusion model for sequential recommendation. Thanks to the diffusion model's inherent capability of distribution generation, we are able to model item's multi-latent aspects and user's diverse intentions as distributions. 
    \item \textcolor{black}{We devise a diffusion model, which can inherently fuse the target item embedding for historical interacted item representation generation.
    Thus, user's current preference and intentions could be introduced as an auxiliary signal for better user understanding. Besides, some uncertainty could also be injected into modeling training process to enhance model robustness.}
    \item We conduct experiments on four real-world datasets to verify the effectiveness of \baby. The results show that \baby outperforms nine existing strong baselines by a large margin. Extensive ablation studies and empirical analysis also elucidate the effectiveness of our proposed components. 
\end{itemize}

The rest of this paper is organized as follows: In Section~\ref{sec:relwork}, we will briefly introduce the relevant works to our method including some typical models for sequential recommendation, multi-interesting modeling, and representative methods specialized in distribution representation and uncertainty injection. In Section~\ref{sec:methodology}, the background knowledge about the diffusion model will be illustrated first. Then, we will present a whole pipeline of \baby so that the readers could obtain a general impression to our method. Subsequently, the key modules and the algorithm in this method, \eg diffusion and reverse phases, the architecture of \textit{Approximator}, will also be elucidated in detail. Section~\ref{sec:exp} will introduce the evaluated public datasets and metrics, experiments setting, and the overall performance of our method compared to other baselines. Additionally, some in-depth analysis \eg training efficiency and convergence, ablation studies, performance on head and long-tail items, and uncertainty visualization, will also be given to exhibit the effectiveness of our method. Finally, we will conclude this work and discuss future research directions in Section~\ref{sec:conclusion}. 

\section{Related Work}
\label{sec:relwork}

We first briefly review the representative methods for sequential recommendation. We then review the related studies on multi-interest modeling and distribution representation generation, the two major areas that are most related to our work. Lastly, we introduce the development of diffusion models in different research areas. 

\subsection{Sequential Recommendation} 

Markov Chain played a crucial role in sequential recommendation in the early stage~\cite{zimdars2013using, shani2005mdp, rendle2010factorizing}, for modeling a user's interaction behavior as a Markov Decision Process (MDP). 
Although Markov chains achieved remarkable success for sequential recommendation in the early years, the strong assumption that the next clicked item only relies on the previous one heavily confined the representation ability of this method.
In deep learning era, many deep sequential neural networks, \eg GRU, LSTM, CNN, and further variants~\cite{hidasi2015session, xiao2019hierarchical}, emerged and delivered promising performance on this task. GRU4REC~\cite{hidasi2015session} is a pioneering work that stacks multiple GRU layers for sequential information modeling and applies a session-parallel mini-batches training strategy for next item recommendation. Caser~\cite{tang2018personalized} models recent K adjacent items as an "image" then apply convolution operation to capture the sequential patterns for next-item recommendation.
Except that, other advancements of sequence modeling, \eg attention mechanisms~\cite{kang2018self, tang2018personalized, sun2019bert4rec} and memory network~\cite{chen2018sequential, tan2021dynamic}, also demonstrated their effectiveness with state-of-the-art performance in this area. SASRec~\cite{kang2018self} and BERT4Rec~\cite{sun2019bert4rec} are two representative works that apply self-attention for sequential recommendation. To be specific, SASRec regards the next-item prediction as an auto-regressive task and utilizes a uni-direction mask self-attention to capture the previous item information for next item prediction. In contrast, BERT4Rec argues that future information will also facilitate the current item predictions. Therefore, it models the sequential recommendation as a cloze task then applies bi-direction self-attention to introduce the future information for recommendation. 
Moreover, Chen \etal~\cite{chen2018sequential} devised a memory-augmented neural network (MANN) with collaborative filtering which contains item-level and future-level specifications for sequential recommendation. On top of that, Tan \etal~\cite{tan2021dynamic} proposed a novel dynamic memory-based attention network (DMAN), which truncates long sequences into several sequence fragments and applies a memory network in a dynamic fashion to extract users' both short-term and long-term interests for recommendation. 

Attributed to information propagation and aggregation, graph neural network is more effective in incorporating high-order relationships in a sequence. Hence it was applied for sequential recommendation~\cite{fan2021continuous, guo2021gcn, ding2021leveraging}. Wu \etal~\cite{wu2019session} proposed SR-GNN, which is the first exploration that introduces a graph neural network to model explicit relationships between items for next-item recommendation.
Additionally, Fan \etal~\cite{fan2021continuous} elaborated to model time interval information and collaborate filter signal with graph neural network for sequential recommendation. Ding \etal ~\cite{ding2021leveraging} constructed the user-item interaction graph and the item-item transition graph to extract more complicated collaborative filtering signals and item transitional patterns for user and item embedding enhancement and sequential recommendation.

\subsection{Multi-Interest Modeling} 
As users' interests are dynamic and diverse, multi-interest modeling with soft-attention mechanism and dynamic routing became prominent for sequential recommendation. For example, Zhuo \etal~\cite{zhou2018deep} believes utilizing a single fixed vector can not fully express the user's diversity preference, thus, a deep interest network (DIN) with a local activation unit was designed to adaptively learn user's interest representation from his/her historical interaction sequence. As such, the expressive capability of the model could be enhanced to fully express users' diverse interests. Besides, MIND~\cite{li2019multi} designed a multi-interest extractor layer via the capsule dynamic routing mechanism for user's multi-interest extraction and sequence recommendation. On top of that, Cen \etal further proposed ComiRec, which leverages two modules \ie dynamic routing and attention mechanism respectively for multi-interests modeling and recommendation.   
More recent works, \eg Re4~\cite{zhang2022re4}, TiMiRec~\cite{wang2022target} strive to introduce auxiliary loss functions for target interest distillation and distinction. 
MGNM~\cite{tian2022multi} combines graph convolution operation with dynamic routing for more precisely multi-level interest learning. 
Considering user's preference evolution process may be uncertain and volatile, SPLIT~\cite{shao2022sequential} applies reinforcement learning to decompose independent preference from user's historical behavior sequence for multi-interest modeling and sequential recommendation.
However, all these solutions require to pre-define a proper number of interests. This process may be heuristic and time-consuming. Additionally, the fixed interest number also confines the model from learning complicated transition patterns in a flexible manner, leading the model to achieve sub-optimal performance.

\subsection{Representation Distribution and Uncertainty Modeling}
As aforementioned, user's preference is uncertain and diverse in practical scenarios, thus, a unified representation vector is insufficient to model user's dynamic preference in an adaptive manner. 
VAE, with its ability to model probabilistic latent variables, has been used for generating representation distributions and injecting uncertainty in sequential recommendation.
For example, Liang \etal proposed Multi-VAE~\cite{liang2018variational}, the pioneering work that applied VAE to collaborative filtering for recommendation. Sachdeva \etal combined recurrent neural network with VAE to capture temporal patterns for sequential recommendation~\cite{sachdeva2019sequential}. Considering the posterior distribution approximation hurts the representation ability of conventional VAE methods, ACVAE introduced the Adversarial variational Bayes (AVB) framework to enhance the representation of latent variables~\cite{xie2021adversarial}. Wang \etal further incorporated contrastive learning into the VAE paradigm to alleviate the representation degeneration problem~\cite{wang2020kerl}. Different from these VAE methods, Fan \etal~\cite{fan2021modeling} proposed a Distribution-based Transform, which models the item representation as Elliptical Gaussian distributions where the conventional unified item embedding is served as means to reveal user's basic interests, besides a stochastic vector is also introduced for covariance representation and uncertainty injection. Moreover, the authors utilized the Wasserstein Distance as a loss function to optimize the distance between sequential distribution representation and item representation distribution. On that basis, Fan \etal~\cite{fan2022sequential} further proposed a novel stochastic self-attention (STOSA) model very recently, which replaces the inner product between any of two items in self-attention module as Wasserstein Distance for correlation measuring and collaborative transitivity capture.
Nevertheless, VAE methods have the ability to model latent variables as distributions, they struggle with representation degeneration and collapse issues. 

\subsection{Diffusion Models} 
Motivated by non-equilibrium thermodynamics~\cite{sohl2015deep}, Diffusion model have shown its great potential in computer vision~\cite{cai2020learning, ho2022cascaded, ho2022video, ramesh2022hierarchical}, natural language processing~\cite{li2022diffusion, gong2022diffuseq, he2022diffusionbert} and other areas~\cite{ chen2020wavegrad}.
For instance, Super-Resolution via Repeated Refinement (SR3)~\cite{saharia2022image} models the super-resolution task as a stochastic, iterative denoising process. Then utilizing Denoising Diffusion Probabilistic Models (DDPM) for conditional image generation. DALLE-2~\cite{ramesh2022hierarchical} applies a two-stage modeling method for text-conditional image generation. To be specific, it first trains a text encoder and image encoder with CLIP strategy for text representation and image representation alignment. Then, given any text description, we could obtain the representation via the well-trained text encoder, which will be further fed into an autoregressive or diffusion prior for image representation generation. Conditioned on this image representation, a diffusion-based decoder is applied to produce the final image. In the NLP area, Diffusion-LM~\cite{li2022diffusion} is the first effort that adapted continuous diffusion to instantiate the idea of fine-grained control on NLP-oriented tasks. 
Following Diffusion-LM, DiffuSeq made the extension to support more general sequence-to-sequence tasks~\cite{gong2022diffuseq}. Specifically, in training or diffusion process, given source and target text pairs, DiffuSeq first corrupts the target text sequence into Gaussian noises. Then, based on the source text sequence and approximator model to reconstruct the target text sequence. In reverse or inference process, given any pure Gaussian noise and source text, we could reverse the target text iteratively from the noise via the well-trained approximator.   
Although Diffusion models became ubiquitous in other domains, adapting the model to recommender systems remains under-explored. We believe Diffusion model is a good fit to SR, and make the first attempt in this paper.  

\section{Preliminary}
\label{sec:preliminary}

Before presenting our model, we first briefly introduce the standard diffusion model as preliminary knowledge. Endowed with the property of latent variable modeling, diffusion model could generate continuous diversity distribution representation in the diffusion and reverse phases. 

In \textbf{\textit{diffusion phase}}, the diffusion model incrementally corrupts the original representation $\mathbf{x}_0$ into a Gaussian noise $\mathbf{x}_t$ via a Markov Chain (\ie $\mathbf{x}_0 \rightarrow \mathbf{x}_1 \rightarrow,\ldots,\rightarrow, \mathbf{x}_{t-1} \rightarrow \mathbf{x}_t$). This process can be formalized as follows:

\begin{equation}
    q(\mathbf{x}_t|\mathbf{x}_{t-1})=\mathcal{N}\left(\mathbf{x}_t;\sqrt{1-\beta_t}\mathbf{x}_{t-1}, \beta_t\mathbf{I}\right)
    \label{eq:diffu_step}
\end{equation}

where $\mathcal{N}(x;\mu,\sigma^2)$ is a Gaussian distribution with a mean $\mu=\sqrt{1-\beta_t}$ and variance $\sigma^2=\beta_t$, $\mathbf{x}_t$ is sampled from this Gaussian distribution, $\beta_t$ controls the noise added at the $t$-th diffusion step and $\mathbf{I}$ is an identity matrix. 

The value of $\beta_t$ is generated from a pre-defined noise schedule $\beta$ which arranges how much noise is injected at which step. The common noise schedule includes square-root~\cite{li2022diffusion}, cosine~\cite{ho2020denoisingdiff}, and linear~\cite{nichol2021improved}. According to to~\cite{ho2020denoisingdiff}, at an arbitrary diffusion step $t$, we can derive $\mathbf{x}_t$ conditioned on the input $\mathbf{x}_0$ in a straightforward way following the Markov chain process. Then Equation~\ref{eq:diffu_step} can be rewritten as follows:

\begin{align}
    q(\mathbf{x}_t|\mathbf{x}_0) &= \mathcal{N} \left(\mathbf{x}_t;\sqrt{\overline{\alpha}_t}\mathbf{x}_0,(1-\overline{\alpha}_t)\mathbf{I}\right)
    \label{eq:diffu}\\
    \overline{\alpha}_t&=\prod_{s=1}^t\alpha_s,~~\alpha_s=1-\beta_s
     \label{eq:alpha_t}
\end{align}

\begin{figure*}[t]
\centerline{\includegraphics[width=\textwidth]{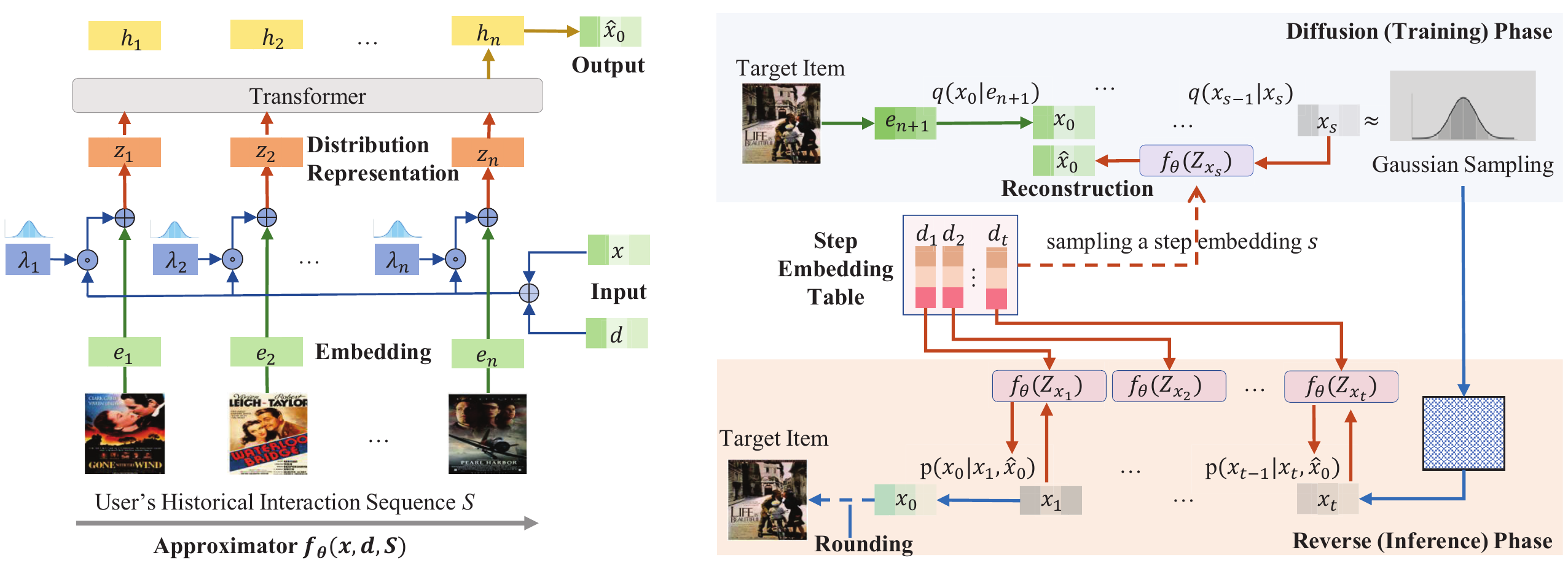}}
\caption{Architecture of \baby. The figure on the left is the \textit{Approximator}, a Transformer backbone for target item representation reconstruction. The two figures on the right illustrate the diffusion phase and the reverse phase, respectively.}
\label{fig:overall_structure}
\end{figure*}

In \textbf{\textit{reverse phase}}, a standard Gaussian representation $\mathbf{x}_t$ is denoised to approximate the real representation $\mathbf{x}_0$ (\ie $\mathbf{x}_t\rightarrow \mathbf{x}_{t-1}\rightarrow,..., \mathbf{x_1}, \rightarrow, \mathbf{x}_0$) in an iterative manner. Specifically, given the current denoised representation $\mathbf{x}_{s}$, the next representation $\mathbf{x}_{s-1}$ after one-step reverse is calculated as follows:

\begin{align}
    p(\mathbf{x}_{s-1}|\mathbf{x}_s,\mathbf{x}_0)&=\mathcal{N}\left(\mathbf{x}_{s-1};\Tilde{\mathbf{\mu}}_s(\mathbf{x}_s,\mathbf{x}_0),\Tilde{\beta}_s\mathbf{I}\right)\label{eq:reverse}\\
\Tilde{\mathbf{\mu}}_s(\mathbf{x}_s,\mathbf{x}_0)&=\frac{\sqrt{\overline{\alpha}_{s-1}}\beta_s}{1-\overline{\alpha}_s}\mathbf{x}_0+\frac{\sqrt{\alpha_s}(1-\overline{\alpha}_{s-1})}{1-\overline{\alpha}_s}\mathbf{x}_s \\ \Tilde{\beta_s}&=\frac{1-\overline{\alpha}_{s-1}}{1-\overline{\alpha}_s}\beta_s 
\end{align}

\textcolor{black}{However, in the reverse phase, the $\mathbf{x}_0$ is unknown, a deep neural network $f_{\theta}(\cdot)$(\eg Transformer~\cite{vaswani2017attention} or U-Net~\cite{ronneberger2015u}) is generally used for estimating $\mathbf{x}_0$. Given the original representation $\mathbf{x}_0$ and the schedule $\beta$, $f_{\theta}(\cdot)$ is trained to minimize the following variational lower bound (VLB)~\cite{sohl2015deep}. Below is the derivation of the VLB, we include it here for completeness:}


\textcolor{black}{
\begin{align}
\label{eq:loss_diffusion}
\mathcal{L}_{vlb}&=\mathbb{E}_q\left[ -\textrm{log} \frac{p_{\theta}(\mathbf{x}_{0:t})}{q(\mathbf{x}_{1:t}|\mathbf{x}_0)}\right]\\\nonumber
&=\mathbb{E}_q\left[-\textrm{log} p(\mathbf{x}_t) - \sum_{t\geq 1} \textrm{log} \frac{p_{\theta}(\mathbf{x}_{t-1}|\mathbf{x}_t)}{q(\mathbf{x}_{t}|\mathbf{x}_{t-1})}\right]\\\nonumber
&=\mathbb{E}_q\left[-\textrm{log} p(\mathbf{x}_t) - \sum_{t>1} \textrm{log} \frac{p_{\theta}(\mathbf{x}_{t-1}|\mathbf{x}_t)}{q(\mathbf{x}_{t}|\mathbf{x}_{t-1})}-\textrm{log} \frac{p_{\theta}(\mathbf{x}_0|\mathbf{x}_1)}{q(\mathbf{x}_1|\mathbf{x}_0)} \right]\\\nonumber
&=\mathbb{E}_q\left[-\textrm{log} p(\mathbf{x}_t) - \sum_{t> 1} \textrm{log} \frac{p_{\theta}(\mathbf{x}_{t-1}|\mathbf{x}_t)}{q(\mathbf{x}_{t-1}|\mathbf{x}_{t},\mathbf{x}_0)} \cdot \frac{q(\mathbf{x}_{t-1}|\mathbf{x}_0)}{q(\mathbf{x}_t|\mathbf{x}_0)} -\textrm{log} \frac{p_{\theta}(\mathbf{x}_0|\mathbf{x}_1)}{q(\mathbf{x}_1|\mathbf{x}_0)} \right]\\\nonumber
&=\mathbb{E}_q\left[-\textrm{log} \frac{p(\mathbf{x}_t)}{q(\mathbf{x}_t|\mathbf{x}_0)} - \sum_{t>1} \textrm{log} \frac{p_{\theta}(\mathbf{x}_{t-1}|\mathbf{x}_t)}{q(\mathbf{x}_{t-1}|\mathbf{x}_t,\mathbf{x}_0)}- \textrm{log} p_{\theta}(\mathbf{x}_0|\mathbf{x}_1)\right]\\\nonumber
&= \mathbb{E}_q \left[ \underbrace{ D_{KL}\left(q(\mathbf{x}_t|\mathbf{x}_0)||p(\mathbf{x}_t)\right)}_{L_t}
+\underbrace{\sum_{t>1} D_{KL}(q(\mathbf{x}_{t-1}|\mathbf{x}_t,\mathbf{x}_0)||p_{\theta}(\mathbf{x}_{t-1}|\mathbf{x}_t))}_{L_{t-1}}
-\underbrace{\textrm{log} p_{\theta}(\mathbf{x}_0|\mathbf{x}_1)}_{L_0}\right]
\end{align}
}

where $p(\mathbf{x}_t)=\mathcal{N}(\mathbf{x}_t;\mathbf{0},\mathbf{I})$, and $D_{KL}(\cdot)$ is KL divergence measure. 

Here, we organize this loss \textcolor{black}{(ref.~Equation~\ref{eq:loss_diffusion})} into three parts: $L_t, L_{t-1}$ and $L_0$. Part $L_t$ works to push the $q(\mathbf{x}_t|\mathbf{x}_0)$ close to a standard Gaussian distribution. Part $L_{t-1}$ works to minimize the KL divergence between the forward process posterior and $p_\theta(\mathbf{x}_{t-1}|\mathbf{x}_t)$ (generated by the deep neural network) for reverse process. The last part $L_0$ utilizes negative log-likelihood for final prediction. Note that this VLB loss could easily lead to unstable model training. 
\textcolor{black}{To alleviate this issue, following~\cite{ho2020denoisingdiff}, we could expand and reweight each KL-divergence term in VLB with a specific parameterization to obtain a simple mean-squared error loss as follows:}

\begin{equation}
    \mathcal{L}_{simple} = \mathbb{E}_{t,\mathbf{x}_0,\mathbf{\epsilon}}\left[||\mathbf{\epsilon}-\mathbf{\epsilon_{\theta}}(\sqrt{\overline{\alpha}_t}\mathbf{x}_0+\sqrt{1-\overline{\alpha}_t}\mathbf{\epsilon}, t)||^2\right]
    \label{eq:obj_simple}
\end{equation}

where $\mathbf{\epsilon}\sim \mathcal{N}(\mathbf{0}, \mathbf{I})$ is sampled from a standard Gaussian distribution for noise injection and diffusion, and $\mathbf{\epsilon_\theta}(\cdot)$ works as a \textbf{\textit{Approximator}} that can be instantiated by a deep neural network \eg Transformer or U-Net.

\section{Methodology}
\label{sec:methodology}

We first formally define the task of sequential recommendation, then we introduce the technical detail of our proposed \baby model including diffusion and revers process for model training and inference respectively, approximator for sequence modeling and target item representation reconstruction, rounding operation for target item index projection, and loss function for model optimization. 

\paratitle{Sequential Recommendation.} Given a set of items $\mathcal{I}$, a set of users $\mathcal{U}$, and their interaction records, we can organize each user's interactions chronologically as a sequence $\mathcal{S}=[i_1^u,i_2^u,...,i_{n-1}^u,i_n^u]$. Here, $u\in \mathcal{U}$, $i_j^u\in \mathcal{I}$ denotes the $j$-th item interacted with user $u$. $n$ is the sequence length and is also the total number of interactions from user $u$. The goal of sequential recommendation is to generate a \textcolor{black}{ranked} list of items as the predicted candidates for next item that will be preferred by the user $u$.

As its name suggests, \baby is built upon the diffusion model. The overall architecture of \baby is shown in Figure~\ref{fig:overall_structure}, with three major parts: 1) Approximator for target item representation reconstruction; 2) Diffusion (Training) process to incorporate the guidance of target item and inject noise for robust approximator learning; and 3) Reverse (Inference) phase for target item prediction.

\subsection{Overview}
\label{sec:overview}

Refer to Figure~\ref{fig:overall_structure}, at first, we consider a static item embedding $\mathbf{e}_j$ as the semantic encoding of the intrinsic latent aspects covered by item $i_j$. Following the diffusion model described in Section~\ref{sec:preliminary}, we plan to perform reverse to recover the target item from the historical sequence $\mathcal{S}$. Hence, we add noise into target item embedding $\mathbf{e}_{n+1}$ through the diffusion process. Recall that the diffusion process involves a series of samplings from Gaussian distributions (ref.~Equation~\ref{eq:diffu_step}). We treat the noised target item representation $\mathbf{x}_s$ undergo $s$-diffusion steps as the distribution representation sampled from $q(\cdot)$. Similar to adversarial learning~\cite{goodfellow2014explaining}, the injected noise can help us to derive sharper item embeddings, thus $\mathbf{x}_s$ can still reflect the information of multiple latent aspects of the item. Then, we use $\mathbf{x}_s$ to adjust the representation of each historical item in $\mathcal{S}$, such that the guidance of target item is exploited as auxiliary semantic signals. Afterwards, the resultant representations $\mathbf{Z}_{x_s}=[\mathbf{z}_1,\mathbf{z}_2,\cdots,\mathbf{z}_n]$ are fed into approximator $f_{\theta}(\cdot)$.
That is, the model training is performed to reinforce the reconstructed $\mathbf{\hat{x}}_0$ from the approximator close to target item embedding $\mathbf{e}_{n+1}$. 

As to the reverse phase, we firstly sample the noised target item representation $\mathbf{x}_t$ from a standard Gaussian distribution, \ie $\mathbf{x}_t\sim \mathcal{N}(\mathbf{0}, \mathbf{I})$. Similar to the diffusion process, the resultant
representations $\mathbf{Z}_{x_t}$ adjusted by using $\mathbf{x}_t$ are fed into the well-trained approximator $f_{\theta}(\cdot)$ for $\mathbf{\hat{x}}_0$ estimate. Following Equation~\ref{eq:reverse}, the estimated $\mathbf{\hat{x}}_0$ and $\mathbf{x}_t$ are used to reverse $\mathbf{x}_{t-1}$ via $p(\cdot)$. This process continues iteratively until $\mathbf{x}_0$ is arrived. Note that the reverse phase is also stochastic, reaching the purpose of modeling uncertainty in user behavior. Then we utilize a rounding function to map the reversed continuous representation $\mathbf{x}_0$  into discrete candidate item indexes for target item $\hat{i}_{n+1}$ prediction.


Formally, the above diffusion, reverse process and rounding operation can be formulated as follows:

\begin{itemize}
    \item \textbf{Diffusion:}
    \begin{equation}
        \begin{split}
            & \mathbf{x}_s \leftarrow q(\mathbf{x}_s|\mathbf{x}_0,s)\\
            & \mathbf{\hat{x}}_0 = f_{\theta}(\mathbf{Z}_{x_s})
        \end{split}
        \label{eq:diffu-process}
    \end{equation}
    \item \textbf{Reverse:}
    \begin{equation}
        \begin{split}
            & \mathbf{\hat{x}}_0= f_{\theta}(\mathbf{Z}_{x_t})\\
            & \mathbf{x}_{t-1} \leftarrow p(\mathbf{x}_{t-1}|\mathbf{\hat{x}}_0,\mathbf{x}_t)
        \end{split}
    \end{equation}
    \item \textbf{Rounding:}
    \begin{equation}
        \hat{i}_{n+1} \leftarrow \textrm{Rounding}(\mathbf{x}_0)
    \end{equation}
\end{itemize}


\subsection{Approximator}
\label{sec:Approximator}


\textcolor{black}{Instead of using U-Net in CV field~\cite{rombach2022high, ramesh2022hierarchical}, following~\cite{gong2022diffuseq, li2022diffusion}, we utilize Transformer as the backbone of Approximator $f_{\theta}(\cdot)$ to generate $\mathbf{\hat{x}}_0$ in diffusion and reverse phases, since Transformer has been well proven to be effective for sequential dependencies modeling and it is ubiquitous in NLP and recommendation tasks.}


\begin{align}   
     \mathbf{\hat{x}}_0 &= f_{\theta}(\mathbf{Z}_x) = \textrm{Transformer}\left([\mathbf{z}_0,\mathbf{z}_1,..,\mathbf{z}_n]\right)\\
    \mathbf{z}_i &= \mathbf{e}_i + \mathbf{\lambda}_i \odot (\mathbf{x}+\mathbf{d})
    \label{eq:approximator}
\end{align}

where symbol $\odot$ is the element-wise product, and $\mathbf{d}$ is the embedding of the corresponding diffusion/reverse step created by following~\cite{vaswani2017attention}. Through step embeddings, the specific information of each diffusion and reverse step could be introduce into the model. As discussed earlier, in diffusion phase, $\mathbf{x}$ is the noised target item embedding $\mathbf{x}_s$ (more discussion in Section~\ref{sec:diffusion}). And in reverse process, $\mathbf{x}$ is the reversed target item representation, \ie $\mathbf{x}=\mathbf{x}_s~(s=t,t-1,\ldots, 2, 1)$. In Equation~\ref{eq:approximator}, $\mathbf{\lambda_i}$ is sampled from a Gaussian distribution, \ie $\lambda_i \sim \mathcal{N}(\mathbf{\delta}, \mathbf{\delta})$, $\delta$ is a hyperparameter which defines both the mean and variance. For diffusion phase,  $\mathbf{\lambda}_i$ controls the extent of noise injection. On the other hand, this setting introduces uncertainty to 
model the user's interest evolution for reverse phase. That is, the importance of each latent aspect of a historical item can be adjusted iteratively along the reverse steps in a user-aware manner.


We keep the same as the standard Transformer architecture and configuration~\cite{vaswani2017attention}, \ie the multi-head self-attention, feed-forward network with ReLU activation function, layer normalization, dropout strategy, and residual connection are utilized. To enhance the representation ability of models, we also stack multi-blocks Transformer. Finally, we gather the representation $\mathbf{h}_{n}$ of item $i_n^u$ generated by the last layer as  $\mathbf{\hat{x}}_0$. 

\subsection{Diffusion Phase}
\label{sec:diffusion}

The key part of the diffusion phase is to design a noise schedule $\beta_s$ to arrange the noise injection in step $s$ for uncertainty modeling. Following~\cite{sohl2015deep, hoogeboom2021argmax}, a uniform noise schedule was explored for the discrete domain. \textcolor{black}{In this paper, we utilize a truncated linear schedule for $\beta_s$ generation (ref. Figure~\ref{fig:schedule} and Figure~\ref{fig:alpha_t} for more detailed discussion) which can be formalized as follows:}


\textcolor{black}{
\begin{equation}
\label{eq:schedule}
\beta_s = \begin{cases}
\frac{a}{t}s+\frac{b}{s}, & \beta_s \leq \tau \\
\frac{1}{10}(\frac{a}{t}s+\frac{b}{s}), &\text{others} \\
\end{cases}
\end{equation}
}

where $a,b$ are hyperparameters, which control the range of $\beta$. If $\beta_s>\tau$, we define $\beta_s=0.1\beta_s$. $\tau$ is a truncated threshold, we set $\tau=1$ in our paper. Denoting the maximum diffusion step as $t$, in the training process, we randomly sample a diffusion step $s$ for each target item, \ie. $s=\lfloor s' \rfloor, s'\sim U(0, t)$. Following Equation~\ref{eq:diffu-process}, we can generate $\mathbf{x}_s$ via 
$q(\mathbf{x}_s|\mathbf{x}_0)=\mathcal{N}(\mathbf{x}_s;\sqrt{\overline{\alpha}_t}\mathbf{x}_0,(1-\overline{\alpha}_s)\mathbf{I})$. We choose to derive $\mathbf{x}_0$ with one step diffusion from the target item embedding: \ie $q(\mathbf{x}_0|\mathbf{e}_{n+1})=\mathcal{N}(\mathbf{x}_0;\sqrt{\alpha_0}\mathbf{e}_{n+1},(1-\alpha_0)\mathbf{I})$. Actually, based on the reparameter trick (\ie $\mathbf{\epsilon}\sim \mathcal{N}(\mathbf{0}, \mathbf{I})$), we could generate $\mathbf{x}_s$ as follows:
\begin{align}
     \mathbf{x}_s = \sqrt{\overline{\alpha}_s}\mathbf{x}_0+\sqrt{1-\overline{\alpha}_s}\mathbf{\epsilon}
\end{align}

\RestyleAlgo{ruled}
\begin{algorithm}[t]
	\renewcommand{\algorithmicrequire}{\textbf{Input:}}
	\renewcommand{\algorithmicensure}{\textbf{Output:}}
	\caption{Reverse or Inference}
	\label{alg:revision}
	\begin{algorithmic}[1]
            \STATE \textbf{Input:}
            \STATE \quad Historical Sequence: $(i_1,\cdots, i_n)$\;
            \STATE \quad Target item Representation $\mathbf{x}_t$: $\mathbf{x}_t\sim \mathcal{N}(\mathbf{0}, \mathbf{I})$\; 
            \STATE \quad Total Reverse Steps: $t$\;
            \STATE \quad Schedule $\beta$: \textit{Linear}\;
            \STATE \quad Hyperparameter $\delta$ (mean and variance): $\lambda$ sampling\;
            \STATE \quad \textit{Approximator}: $f_\theta(\mathbf{x})$\;
            \STATE \textbf{Output:}
            \STATE \quad \quad Predicted Target Item: $i_{n+1}$;
            \BlankLine
        $s = t$
		\WHILE{$s > 0$}
            \STATE$(\mathbf{z}_1,..., \mathbf{z}_{n})$ $\leftarrow \left( \mathbf{e}_1+\lambda_1 (\mathbf{x}_s+\mathbf{d}_s),..., \mathbf{e}_{n}+\lambda_{n} (\mathbf{x}_s+\mathbf{d}_s) \right)$, 
            $\lambda_i\sim \mathcal{N}(\delta, \delta)$
            ; \tcp{Distribution Representation}
            \STATE $\mathbf{\hat{x}}_0$  $\leftarrow f_{\theta} (\mathbf{z}_1,..., \mathbf{z}_{n})$; \tcp{Reconstruction}
            \STATE $\mathbf{x}_{s-1}$ $\leftarrow \Tilde{\mu}(\mathbf{\hat{x}}_0, \mathbf{x}_s) + \Tilde{\beta}_t*\epsilon'$, $(\epsilon' \sim \mathcal{N}(\mathbf{0},\mathbf{I}))$; \tcp{reverse}
            \STATE $s = s - 1$; \tcp{Iteration}
            \ENDWHILE
        \STATE $i_{n+1}$  $\leftarrow \textrm{Rounding}(\mathbf{x}_0)$;
	\end{algorithmic}  
\label{alg:reverse}
\end{algorithm}

\subsection{Reverse Phase}
\label{sec:reverse}
As aforementioned, we aim to recover the target item representation $\mathbf{x}_0$ iteratively from a pure Gaussian noise $\mathbf{x}_t$, in the reverse phase. However, this process will be intractable as $\mathbf{x}_0$ is required at each reverse step. Consequently, we use the well-trained approximator to generate $\mathbf{\hat{x}}_0$ for $\mathbf{x}_0$ estimation, \ie $\mathbf{x}_0=\mathbf{\hat{x}}_0$. Following Equation~\ref{eq:reverse}, the reverse step is performed after applying reparameter trick:
\begin{align}
    \mathbf{\hat{x}}_0 &= f_{\theta}(\mathbf{Z}_{x_t})\\
    \mathbf{x}_{t-1} &= \Tilde{\mathbf{\mu}}_t(\mathbf{x}_t,\mathbf{\hat{x}}_0)+\Tilde{\beta}_t\mathbf{\epsilon'}
    \label{eq:rev}
\end{align}

where $\Tilde{\mathbf{\mu}}_t(\mathbf{x}_t,\mathbf{\hat{x}}_0)=\frac{\sqrt{\overline{\alpha}_{t-1}}\beta_t}{1-\overline{\alpha}_t}\mathbf{\hat{x}}_0+\frac{\sqrt{\alpha_t}(1-\overline{\alpha}_{t-1})}{1-\overline{\alpha}_t}\mathbf{x}_t$ and $\Tilde{\beta_t}=\frac{1-\overline{\alpha}_{t-1}}{1-\overline{\alpha}_t}\beta_t$, and $\mathbf{\epsilon'}\sim \mathcal{N}(\mathbf{0}, \mathbf{I})$ for $\mathbf{x}_{t-1}$ generation. Repeat the above process until we arrive at $\mathbf{x}_0$. The reverse phase can be illustrated in Algorithm~\ref{alg:revision}.

\begin{algorithm}[t]
	\renewcommand{\algorithmicrequire}{\textbf{Input:}}
	\renewcommand{\algorithmicensure}{\textbf{Output:}}
	\caption{Diffusion or Training}
	\label{alg:diffusion}
	\begin{algorithmic}[1]
            \STATE \textbf{Input:}
            \STATE \quad Historical Sequence: $(i_1,\cdots, i_{n},i_{n+1})$\;
            \STATE \quad Learning Epochs: $E$\;
            \STATE \quad Maximum Diffusion Steps: $t$\;
            \STATE \quad Schedule $\beta$: \textit{Linear}\;
            \STATE \quad Hyperparameter $\delta$ (mean and variance): $\lambda$ sampling\; 
            \STATE \quad \textit{Approximator}: $f_\theta(\cdot)$\;
            \STATE \textbf{Output:}
            \STATE \quad Well-trained \textit{Approximator}: $f_\theta(\cdot)$\;
            \BlankLine
		\WHILE{$j < E$}
            \STATE $s=\lfloor s' \rfloor, s'\sim U(0, t)$; \tcp{Diffusion Step Sampling}
            \STATE $\mathbf{x}_s$ $\leftarrow q(\mathbf{x}_{s}|\mathbf{x}_{0})$; \tcp{Diffusion}
            \STATE $(\mathbf{z}_1,..., \mathbf{z}_{n})$ $\leftarrow (\mathbf{e}_1+\lambda_1 (\mathbf{x}_s+\mathbf{d}_s),..., \mathbf{e}_{n}+\lambda_{n} (\mathbf{x}_s+\mathbf{d}_s))$, $\lambda_i\sim \mathcal{N}(\delta, \delta)$;\tcp{Distribution Representation}
            \STATE $\mathbf{\hat{x}}_0$ $\leftarrow f_\theta ([\mathbf{z}_1,..., \mathbf{z}_{n}])$; \tcp{Reconstruction}
            \STATE parameter update: $\mathcal{L}_{CE}(\mathbf{\hat{x}}_0,i_{n+1})$;
            \STATE $j = j + 1$;
            \ENDWHILE
	\end{algorithmic}  
\end{algorithm}

\subsection{Loss Function and Rounding}
\label{loss_function}

As the approximator requires the step embedding for reconstruction (ref.~Equatoin~\ref{eq:approximator}), in diffusion phase, we randomly sample a step-index $s$ from a uniform distribution over range $[1,t]$. So as to the reverse phase, the step-index will be from $t$ to $1$, where $t$ is the total steps. Therefore, the corresponding step embedding and sequence item distribution are fed into the approximiator for model learning. 

The standard objective function of diffusion models could be simplified as a mean-squared error (ref.~Equation~\ref{eq:obj_simple}). However, we argue it may be inappropriate in recommendation tasks for two reasons. First, same as NLP, rather than a continuous distribution, the item embedding is static in the latent space. \textcolor{black}{Thus, the mean square error loss is known to be unstable in such a scenario~\cite{dieleman2022continuous, mahabadi2023tess}}. Second, it is more ubiquitous to calculate the relevance between two vectors in terms of inner product for sequential recommendation. Consequently, following~\cite{han2022ssd, mahabadi2023tess, strudel2022self}, in diffusion phase, we instead utilize cross-entropy loss for model optimization as follows:
\begin{align}
 \hat y &=\frac{\textrm{exp}(\mathbf{\hat{x}}_0 \cdot \mathbf{e}_{n+1})}{\sum_{i\in \mathcal{I}}\textrm{exp}(\mathbf{\hat{x}}_0\cdot \mathbf{e}_i)} \label{eq:haty}\\
        \mathcal{L}_{CE} &= \frac{1}{|\mathcal{U}|}\sum_{i\in \mathcal{U}}-\textrm{log}\hat{y}_i
    \label{eq:ce_loss}
\end{align}
where $\mathbf{\hat{x}}_0$ is reconstructed by the Transformer based approximator and symbol $\cdot$ indicates the inner product operation. 

As for the inference phase, we need to map the reversed target item representation $\mathbf{x}_0$ into the discrete item index space for final recommendation. 
\textcolor{black}{Here, we calculate the inner product between the $\mathbf{x}_0$ and all candidate item embeddings $\mathbf{e}_i$. And then, take the item index corresponding to the maximal as the final recommendation result. This can be formalized as follows:}
    \begin{align}
	\mathop{\arg\max}_{i \in \mathcal{I}} \quad \textrm{Rounding}(\mathbf{x}_0)=\mathbf{x}_0\cdot \mathbf{e}_i^T
    \label{eq:rounding}
    \end{align}
Further, we also utilize inner product to rank the candidate items. The diffusion process or model training is in Algorithm~\ref{alg:diffusion}.

\subsection{Discussion}
\label{sec:discussion}

It is noteworthy that attributed to the property of diffusion model the corrupt or reversed target item representation $\mathbf{x}$ will be some uncertainty in diffusion and reverse process. Additionally, we randomly sample a $\lambda$ from Gaussian distribution for item distribution representation generation, this operation will also introduce some stochasticity and uncertainty in model training. We believe the injected uncertainty, \ie the small perturbation to the raw item embedding, will improve the robustness of model and also alleviate the over-fitting problem. To in-depth analyze the effects of this perturbation on the model final performance, following~\cite{he2018adversarial, goodfellow2014explaining}, we implement an adversarial training framework, by adding adversarial perturbations on the embedding vectors for more robust model training and compare the performance of this strategy to our \baby.  

More concretely, we suppose the adversarial perturbations to item embeddings that cause the dramatic change to the final results (a much worse recommendation performance) will be the effective perturbations. Hence, the model is not that robust and is vulnerable to such perturbations. We, thereby, apply adversarial training to improve the model's robustness to those perturbations. To be specific, the process of adversarial training can be transformed into a min-max optimization problem. First, by maximizing the loss function, we aim to find effective perturbations $\Delta$. Afterward, the model is trained to minimize both the original loss $\mathcal{L}(\mathcal{D}|\Theta)$ and the additional loss with adversarial perturbations $\mathcal{L}(\mathcal{D}|\Theta+\Delta)$. The min-max optimization can be formalized as follows:

\begin{align}
\Theta^*, \Delta^* = \mathop{\arg\min}_{\Theta} \mathop{max}_{\Delta, ||\Delta||\leq \epsilon} \mathcal{L}(\mathcal{D}|\Theta)+\gamma\mathcal{L}(\mathcal{D}|\Theta+\Delta) \label{eq:min-max}
\end{align}

where $\epsilon$ regulates the magnitude of the perturbations, $\gamma$ balances the adversarial regularizer and main loss, \ie the cross-entropy loss for target item prediction. $||\cdot||$ denotes the $L_2$ norm. $\Theta$ is the learnable parameters of model.

However, it is intractable to deliver the exact value $\Delta_{adv}=\mathop{\arg\max}_{\Delta,||\Delta||\leq \epsilon}\mathcal{L} (\mathcal{D}|\Theta+\Delta)$ from Equation~\ref{eq:min-max}. As we add the perturbation in a linear manner for adversarial samples construction, following~\cite{goodfellow2014explaining}, we could move towards the gradient direction of the objective function with respect to $\Delta$ when model training to approximate $\Delta_{adv}$ incrementally. This process can be formalized as follows:

\begin{align}
\Delta_{adv}=\epsilon \frac{\Gamma}{||\Gamma||} \quad \textrm{where} \quad \Gamma=\frac{\partial\mathcal{L}(\mathcal{D}|\Theta+\Delta^t)}{\partial\Delta^t} \label{eq:delta}
\end{align}

Denotes $\Delta^t$ as the perturbations in $t$-th training epoch. We initialize $\Delta^0$ as zeros, and define $\Delta^{t+1}=\Delta_{adv}$.

Hence, the min-max optimization function can be rewritten as follows:

\begin{align}
    \mathcal{L}^*(\mathcal{D}|\Theta)=\mathcal{L}(\mathcal{D}|\Theta)+\gamma\mathcal{L}(\mathcal{D}|\Theta+\Delta_{adv})\label{eq:adv_loss}
\end{align}

We use Equation~\ref{eq:adv_loss} as the final loss function for adversarial training. For a fair comparison, we also apply the standard Transformer as a backbone and obtain the last hidden representation $\mathbf{h}_n$ with regard to item $i_n$ as the sequence representation and calculate its inner product with candidate item embeddings to yield the prediction scores for the next-item recommendation. Besides, the other setting, \ie item embedding size, hidden state dimensions, the number of transformer blocks and multi-heads, dropout ratio, etc, keep the same as \baby. The detailed experiment results analysis is presented in Section~\ref{sec:adv}.  

\section{Experiments}
\label{sec:exp}

In this section, we select nine baselines covering three types of mainstream methods in sequential recommendation task, \eg representative sequential recommendation methods, multi-interesting modeling methods, VAE and uncertainty models. We conduct extensive experiments over four real-world datasets for performance evaluation. In detail, we aim to answer the following research questions:   

\begin{itemize}
   \item 
    \textbf{RQ1.} How does the \baby perform compared with state-of-the-art sequential recommendation methods?
    \item 
    \textbf{RQ2.} How does each design choice made in \baby affect its performance?
    \item 
    \textbf{RQ3.} How do different hyperparameter settings affect its performance?
    \item
    \textbf{RQ4.} How does the \baby perform compared with adversarial training?
    \item
    \textbf{RQ5.} How does \baby perform on sequences of varying lengths and items of different popularity?
    \item 
    \textbf{RQ6.} How efficient \textcolor{black}{(\ie time complexity)} is the training and inference  of our proposed model compared with other methods?
    \item
    \textcolor{black}{\textbf{RQ7.} How about the diversity and uncertainty of the final recommendation results of \baby?}
\end{itemize}

\subsection{Datasets and Evaluation Metrics}
\label{datasets}
We use four real-world datasets to validate the efficacy of our \baby. 
All the datasets have been widely used for sequential recommendation. 

\begin{itemize}
    \item 
    \textit{Amazon Beauty} and \textit{Amazon Toys}\footnote{\url{https://cseweb.ucsd.edu/~jmcauley/datasets/amazon_v2/}} are two categories of Amazon review datasets
    , which contains a collection of user-item interactions on Amazon spanning from May 1996 to July 2014. 
    \item 
    \textit{Movielens-1M}\footnote{\url{http://files.grouplens.org/datasets/movielens/ml-1m.zip}} is a widely-used benchmark dataset that includes one million movie ratings from 6000 users on 4000 movies. 
    \item 
    \textit{Steam}\footnote{\url{https://steam.internet.byu.edu/}} is collected from a large online video game distribution platform. The dataset contains more than 40k games with abundant external information (spanning from October 2010 to January 2018), \eg users’ play hours, price, category, media score, publisher and developer information.
\end{itemize}

Following the common data preprocessing method~\cite{kang2018self, sun2019bert4rec, wang2022target}, 
we treat all reviews or ratings as implicit feedback (\ie a user-item interaction) and chronologically organize them by their timestamps. 
Additionally, we filter out unpopular items and inactive users with fewer than 5 related actions. Moreover, we adopt a leave-one-out strategy for performance evaluation. To be specific, for all the datasets, given a sequence $\mathcal{S}=\{i_1,i_2,...,i_n\}$, we utilize the most recent interaction ($i_n$) for testing, the penultimate interaction ($i_{n-1}$) for model validation, and the earlier ones ($\{i_1,i_2,...,i_{n-2}\}$) for model training. The maximum sequence length is set to $200$ for \textit{MovieLens-1M} dataset, and $50$ for the other three datasets. 
The statistics of these datasets are reported in Table~\ref{tab:statistics}. We could find that the average length of sequences and the size of these datasets are very different, covering a broad spectrum of real-world scenarios. 

As for the evaluation metric, we evaluate all models with HR@K (Hit Rate) and NDCG@K (Normalized Discounted Cumulative Gain). We report the experimental results with $K=\{5, 10, 20\}$.
HR@K represents the proportion of the hits recommended among the top-K list. NDCG@K further evaluates the ranking performance by considering the ranking positions of these hits. The NDCG@K is set to $0$ when the rank exceeds K. 
Because the evaluation with sampling may cause inconsistency when the number of negative items is small~\cite{krichene2022sampled}, we rank all candidate items for target item prediction. 



\begin{table}[t]
\caption{Statistics of datasets after preprocessing. Avg\_len means the average length of sequences.}
\small
\setlength\tabcolsep{2pt} 
\begin{center}
\begin{tabular}{lrrrrr}
\hline
\textbf{Dataset}          & \multicolumn{1}{l}{\textbf{\# Sequence}} & \multicolumn{1}{l}{\textbf{\# items}} & \multicolumn{1}{l}{\textbf{\# Actions}} & \multicolumn{1}{l}{\textbf{Avg\_len}} & \multicolumn{1}{l}{\textbf{Sparsity}} \\ \hline
Beauty             &  22,363                                  & 12,101           & 198,502                              & 8.53   & 99.93\%                               \\

Toys & 19,412                                    & 11,924                                 & 167,597                              & 8.63   & 99.93\%                               \\
Movielens-1M              & 6,040                                   & 3,416                                 & 999,611                                & 165.50 & 95.16\%                               \\
Steam                     & 281,428                                         & 13,044                                      & 3,485,022                                       & 12.40 & 99.90\%                                      \\ \hline
\end{tabular}
\label{tab:statistics}
\end{center}
\end{table}

\subsection{Baselines}
\label{sec:eval}

We evaluate \baby against three types of representative sequential recommendation methods, including four conventional SR models, two models designed for multi-interest modeling, and three models dedicated to uncertainty modeling.  

\paratitle{The Conventional Sequential Neural Models}
\begin{itemize}
    \item \textbf{GRU4Rec}\footnote{\url{https://github.com/hidasib/GRU4Rec}}~\cite{hidasi2015session} utilizes GRU to model the sequential behavior of users for recommendation. 
    \item \textbf{Caser}\footnote{\url{https://github.com/graytowne/caser}}~\cite{tang2018personalized} devises horizontal and vertical CNN to exploit user's recent sub-sequence behaviors for recommendation.  
    \item \textbf{SASRec}\footnote{\url{https://github.com/kang205/SASRec}}~\cite{kang2018self} utilizes a single-direction Transformer with a mask encoder to model the implicit correlations between items and it is a competitive benchmark in sequential recommendation task.  
    \item \textbf{BERT4Rec}\footnote{\url{https://github.com/FeiSun/BERT4Rec}}~\cite{sun2019bert4rec} believes the uni-directional architecture (\eg SASRec) is insufficient for users’ behaviors modeling. Consequently, a bidirectional Transformer with cloze task is proposed for sequential recommendation.      
\end{itemize}

\paratitle{Multi-Interest Models} 
\begin{itemize}
    \item \textbf{ComiRec}\footnote{\url{https://github.com/THUDM/ComiRec}}~\cite{cen2020controllable} adopts attention mechanism and dynamic routing for user's multi-interest extraction and recommendation.  
    \item \textbf{TiMiRec}\footnote{\url{https://github.com/THUwangcy/ReChorus/tree/CIKM22}}~\cite{wang2022target} is the most uptodate multi-interest model. Compared with ComiRec, it designs an auxiliary loss function which involves the target item as a supervised signal for interest distribution generation.
\end{itemize}

\paratitle{VAE and Uncertainty Models}
\begin{itemize}
    \item \textbf{SVAE}\footnote{\url{https://github.com/noveens/svae_cf}}~\cite{sachdeva2019sequential} is a pioneer work that combines GRU and variational autoencoder for next item prediction.
    \item \textbf{ACVAE}\footnote{\url{https://github.com/ACVAE/ACVAE-PyTorch}}~\cite{xie2021adversarial} utilizes the Adversarial Variational Bayes (AVB) framework and propose an Adversarial and Contrastive Variational Autoencoder to generate high-quality latent variable representations for sequential recommendation.
    \item \textbf{STOSA}\footnote{\url{https://github.com/zfan20/STOSA}}~\cite{fan2022sequential} develops the vanilla self-attention as a Wasserstein self-attention to model the inner correlation between any of two item representations and also inject some uncertainty into the model training process via a stochastic Gaussian distribution for sequential recommendation.
\end{itemize}

\begin{table*}
\caption{Experimental results (\%) on the four datasets. The best results are in boldface, and the second-best underlined. Symbol $\blacktriangle\%$ is the relative improvement of \baby against the best baseline. * denotes a significant improvement over the best baseline (t-test P<.05).}
\resizebox{\textwidth}{!}{
\begin{tabular}{clccccccccccc}
\hline
Dataset                       & Metric  & GRU4Rec & Caser   & SASRec  & BERT4Rec & ComiRec & TiMiRec & SVAE   & ACVAE   & STOSA   & DiffuRec &  $\blacktriangle\%$       \\ \hline
\multirow{6}{*}{\rotatebox[origin=c]{90}{Beauty}}       & HR@5    & 1.0112  & 1.6188  & 3.2688  & 2.1326   & 2.0495  & 1.9044  & 0.9943 & 2.4672  & \underline{3.5457}  & \textbf{5.5758*}   & 57.26\% \\
                              & HR@10   & 1.9370  & 2.8166  & \underline{6.2648}  & 3.7160   & 4.4545  & 3.3434  & 1.9745 & 3.8832  & 6.2048  & \textbf{7.9068*}   & 26.21\% \\
                              & HR@20   & 3.8531  & 4.4048  & 8.9791  & 5.7922   & 7.6968  & 5.1674  & 3.1552 & 6.1224  & \underline{9.5939}  & \textbf{11.1098*}  & 15.80\% \\
                              & NDCG@5  & 0.6084  & 0.9758  & 2.3989  & 1.3207   & 1.0503  & 1.2438  & 0.6702 & 1.6858  & \underline{2.5554}  & \textbf{4.0047*}   & 56.72\% \\
                              & NDCG@10 & 0.9029  & 1.3602  & \underline{3.2305}  & 1.8291   & 1.8306  & 1.7044  & 0.9863 & 2.1389  & 3.2085  & \textbf{4.7494*}   & 47.02\% \\
                              & NDCG@20 & 1.3804  & 1.7595  & 3.6563  & 2.3541   & 2.6451  & 2.1627  & 1.2867 & 2.7020  & \underline{3.7609}  & \textbf{5.5566*}   & 47.75\% \\ \hline
\multirow{6}{*}{\rotatebox[origin=c]{90}{Toys}} & HR@5    & 1.1009  & 0.9622  & \underline{4.5333}  & 1.9260   & 2.3026  & 1.1631  & 0.9109 & 2.1897  & 4.2236  & \textbf{5.5650*}   & 22.76\% \\
                              & HR@10   & 1.8553  & 1.8317  & 6.5496  & 2.9312   & 4.2901  & 1.8169  & 1.3683 & 3.0749  & \underline{6.9393}  & \textbf{7.4587*}   & 7.48\%  \\
                              & HR@20   & 3.1827  & 2.9500  & 9.2263  & 4.5889   & 6.9357  & 2.7156  & 1.9239 & 4.4061  & \underline{9.5096} & \textbf{9.8417*}   & 3.49\%  \\
                              & NDCG@5  & 0.6983  & 0.5707  & 3.0105  & 1.1630   & 1.1571  & 0.7051  & 0.5580 & 1.5604  & \underline{3.1017}  & \textbf{4.1667*}   & 34.34\% \\
                              & NDCG@10 & 0.9396  & 0.8510  & 3.7533  & 1.4870   & 1.7953  & 0.9123  & 0.7063 & 1.8452  & \underline{3.8806}  & \textbf{4.7724*}   & 22.98\% \\
                              & NDCG@20 & 1.2724  & 1.1293  & 4.3323  & 1.9038   & 2.4631  & 1.1374  & 0.8446 & 2.1814  & \underline{4.3789}  & \textbf{5.3684*}   & 22.60\% \\ \hline
\multirow{6}{*}{\rotatebox[origin=c]{90}{Movielens-1M}}  & HR@5    & 5.1139  & 7.1401  & 9.3812  & 13.6393  & 6.1073  & \underline{16.2176} & 1.4869 & 12.7167 & 7.0495  & \textbf{17.9659*}  & 10.78\% \\
                              & HR@10   & 10.1664 & 13.3792 & 16.8941 & 20.5675  & 12.0406 & \underline{23.7142} & 2.7189 & 19.9313 & 14.3941 & \textbf{26.2647*}  & 10.76\% \\
                              & HR@20   & 18.6995 & 22.5507 & 28.318  & 29.9479  & 21.0094 & \underline{33.2293} & 5.0326 & 28.9722 & 24.9871 & \textbf{36.7870*}  & 10.71\% \\
                              & NDCG@5  & 3.0529  & 4.1550  & 5.3165  & 8.8922   & 3.5214  & \underline{10.8796} & 0.9587 & 8.2287  & 3.7174  & \textbf{12.1150*}  & 11.36\% \\
                              & NDCG@10 & 4.6754  & 6.1400  & 7.7277  & 11.1251  & 5.4076  & \underline{13.3059} & 1.2302 & 10.5417 & 6.0771  & \textbf{14.7909*}  & 11.16\% \\
                              & NDCG@20 & 6.8228  & 8.4304  & 10.5946 & 13.4763  & 7.6502  & \underline{15.7019} & 1.8251 & 12.8210 & 8.7241  & \textbf{17.4386*}  & 11.06\% \\ \hline
\multirow{6}{*}{\rotatebox[origin=c]{90}{Steam}}         & HR@5    & 3.0124  & 3.6053  & 4.7428  & 4.7391   & 2.2872  & \underline{6.0155}  & 3.2384 & 5.5825  & 4.8546  & \textbf{6.6742*}   & 10.95\% \\
                              & HR@10   & 5.4257  & 6.4940  & 8.3763  & 7.9448   & 5.4358  & \underline{9.6697}  & 5.8275 & 9.2783  & 8.5870  & \textbf{10.7520*}  & 11.19\% \\
                              & HR@20   & 9.2319  & 10.9633 & 13.6060 & 12.7332  & 10.3663 & \underline{14.8884} & 7.9753 & 14.4846 & 14.1107 & \textbf{16.6507*}  & 11.84\% \\
                              & NDCG@5  & 1.8293  & 2.1586  & 2.8842  & 2.9708   & 1.0965  & \underline{3.8721}  & 1.8836 & 3.5429  & 2.9220  & \textbf{4.2902*}   & 10.80\% \\
                              & NDCG@10 & 2.6033  & 3.0846  & 4.0489  & 4.0002   & 2.1053  & \underline{5.0446}  & 2.6881 & 4.7290  & 4.1191  & \textbf{5.5981*}   & 10.97\% \\
                              & NDCG@20 & 3.5572  & 4.2073  & 5.3630  & 5.2027   & 3.3434  & \underline{6.3569}  & 3.2323 & 6.0374  & 5.5072  & \textbf{7.0810*}   & 11.39\% \\ \hline
\end{tabular}}
\label{tab:baselines}
\end{table*}


\subsection{Implementation Details}
\label{sec:parameter}
For a fair comparison, we followed~\cite{kang2018self, zhou2022filter, sun2019bert4rec} and utilized the Adam optimizer with the initial learning rate of $0.001$. 
We initialized the parameters of Transformer based on a \textcolor{black}{Xavier} normalization distribution 
and set the block numbers as $4$. We fixed both the embedding dimension and all the hidden state size as $128$ and batch size as $1024$. The dropout rate of turning off neurons in Transformer block and item embedding are $0.1$ and $0.3$ for all datasets. So as to $\mathbf{\lambda}$, we sample these values from a Gaussian distribution with both mean and variance of $0.001$. The total number of reverse steps $t$ is $32$ and we further analyze the efficiency of reverse phase within the scope of $\{2, 4, 8, 16, 32, 64, 128, 256, 512, 1024\}$ (more discussion in Section~\ref{efficiency}). As for the noise schedule $\beta$, we select \textit{truncated linear} schedule for performance comparison (ref.~Equation~\ref{eq:schedule}) and also investigate the effects of \textit{sqrt}, \textit{cosine} and \textit{linear} schedules~\cite{li2022diffusion, nichol2021improved, nichol2021improved} to the performance in Section~\ref{sec:Q2}. For each method, the experiments are conducted five times and the averaged results are reported. We conduct the student $t$-$test$ for statistical significance test. 

\subsection{Overall Comparison (RQ1)}
\label{sec:Q1}

The overall performance of our proposed \baby and the other baselines is presented in Table~\ref{tab:baselines}. Here, we can make the following observations.

\baby consistently outperforms all baselines across the four datasets in terms of all six metrics. In particular, compared with the best baseline, \baby achieves up to $57.26\%$/$56.72\%$, $22.76\%$/$34.34\%$, $10.78\%$/$11.36\%$ and $11.84\%$/$11.39\%$ (HR/NDCG) improvements on \textit{Amazon Beauty}, \textit{Amazon Toys}, \textit{Movielens-1M} and \textit{Steam} datasets, respectively. The significant performance gain suggests that \baby can effectively accommodate the four characteristics (\ie multiple latent aspects of items, multiple interests of users, uncertainty, and guidance of target item) in a unified framework.

Due to the limited capacity of long-term dependency modeling, GRU4Rec and Caser deliver unsatisfied results compared with other baselines across all the datasets.
To address this issue, SASRec endeavors to apply uni-directional Transformer to capture more complicated dependency relations for recommendation. 
Moreover, BERT4Rec believes that future data (\ie the interacted records after current interaction) is also beneficial for sequential recommendation, thus a bi-directional Transformer is utilized instead. The experiment results illustrate that the Transformer remains an efficient and effective model and achieves better performance than GRU4Rec and Caser in most settings.

On the other hand, the multi-interest models like ComiRec and TiMiRec cannot achieve overwhelm superiority against the conventional sequential neural models across all the datasets, especially compared against SASRec and BERT4Rec. Furthermore, on different size datasets, ComiRec and TimiRec perform differently. We speculate it is because the ComiRec pays more attention to the disparity of sequence representations while ignoring the current intention modeling. In contrast, TiMiRec introduces the target item as auxiliary supervised signals for multi-interest distribution generation when model training. Thus, it achieves much better performance than ComiRec on more complicated and long sequence datasets, \eg \textit{Movielens-1M} and \textit{Steam}. 

As a pioneer work of VAE-based solutions for sequential recommendation, SVAE only utilizes GRU for sequential modeling and reports the worst results. ACVAE introduces adversarial learning via AVB framework to enhance the representation ability and acquire a competitive performance in most settings. Additionally, based on Gaussian distribution and Wasserstein distance modeling, STOSA improves the standard self-attention by exploiting dynamic uncertainty injection and distribution representation. In general, the results of STOSA across all the datasets firmly confirm the effectiveness of distribution representation learning and uncertainty injection for sequential recommendation.

\subsection{Ablation Study (RQ2)}
\label{sec:Q2}

To verify the effectiveness of each design choice in \baby, we replace one of them each time to analyze the performance variation. 

\begin{itemize}
    \item \textbf{w GRU}: replaces the Transformer module with a GRU as the \textit{Approximater} and obtain the representation derived at the last step as  $\mathbf{\hat{x}_0}$.  
    \item \textbf{w R*}: replaces the inner product of rounding operation in Equation~\ref{eq:rounding} with the following variant as in~\cite{gong2022diffuseq}:
    \begin{align}
    \frac{2\mathbf{x}_0\cdot \mathbf{e}_i^T}{||\mathbf{x}_0||||\mathbf{e}_i||}-(\frac{\mathbf{x}_0}{||\mathbf{x}_0||}+\frac{\mathbf{e}_i}{||\mathbf{e}_i||})
    \end{align} 
    \item \textbf{w Cosine}: replaces the \textit{truncated linear} schedule with \textit{cosine} schedule as in~\cite{ho2020denoisingdiff} for noise arrangement.
    \item \textbf{w L}: replaces the \textit{truncated linear} schedule with \textit{linear} schedule as in~\cite{nichol2021improved} for noise arrangement.
    \item \textbf{w Sqrt}: replaces the \textit{truncated linear} schedule with \textit{sqrt-root} schedule as in~\cite{li2022diffusion} for noise arrangement.
\end{itemize}

Table~\ref{tab:ablation} shows the ablation results on these four datasets. It is encouraging to note that when we replace the \textit{Approximator} from Transformer backbone to GRU for target item representation reconstruction, \baby still achieves superior performance on \textit{Movielens-1M} and \textit{Steam} datasets while the space and time complexity drop significantly. And this phenomenon is also observed in CV domain that compared with other sophisticated model architectures, a U-Net (a simple Encode-Decoder framework with convolution operation) combined with Diffusion could also achieve extraordinary results. Additionally, we also select the rounding strategy proposed in~\cite{gong2022diffuseq} as an alternative for target item prediction, but \textit{HR}/\textit{NDCG} reduce by at least $68.71\%/74.61\%$, $92.40\%/95.64\%$, $75.08\%/75.93\%$ and $74.12\%/76.17\%$ on \textit{Amazon Beauty}, \textit{Amazon Toys}, \textit{Movielens-1M} and \textit{Steam} datasets, respectively. We believe it is because inner product operation, as most works leveraged, is aligned with objective loss and may be more suitable for sequential recommendation~\cite{kang2018self, sun2019bert4rec, cen2020controllable}. 

\begin{figure}[t]
\centerline{\includegraphics[width=\textwidth]{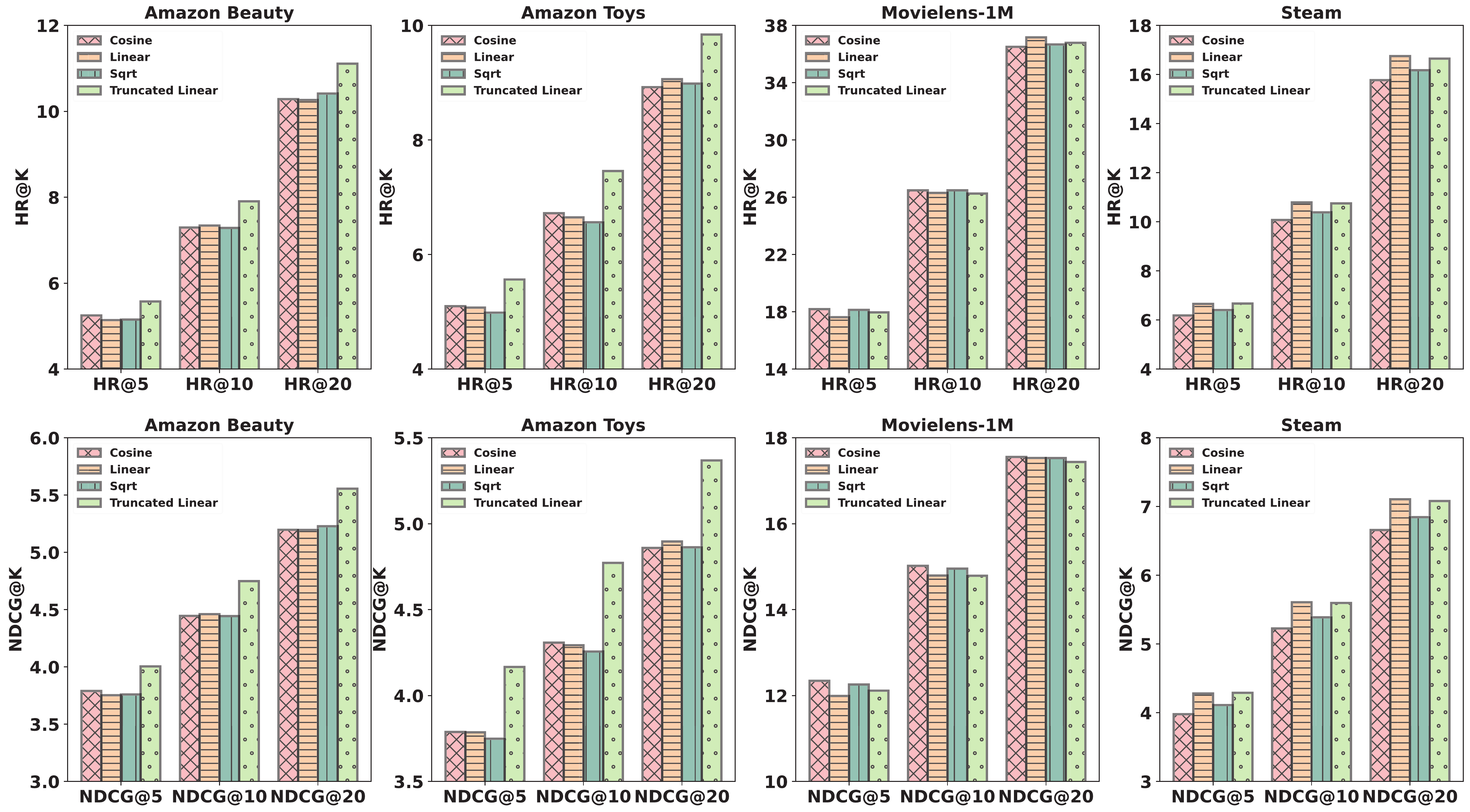}}
\caption{\textcolor{black}{Performance of \baby with different schedule.}}
\label{fig:schedule}
\end{figure}

\begin{figure}[t]
\centerline{\includegraphics[width=0.6\textwidth]{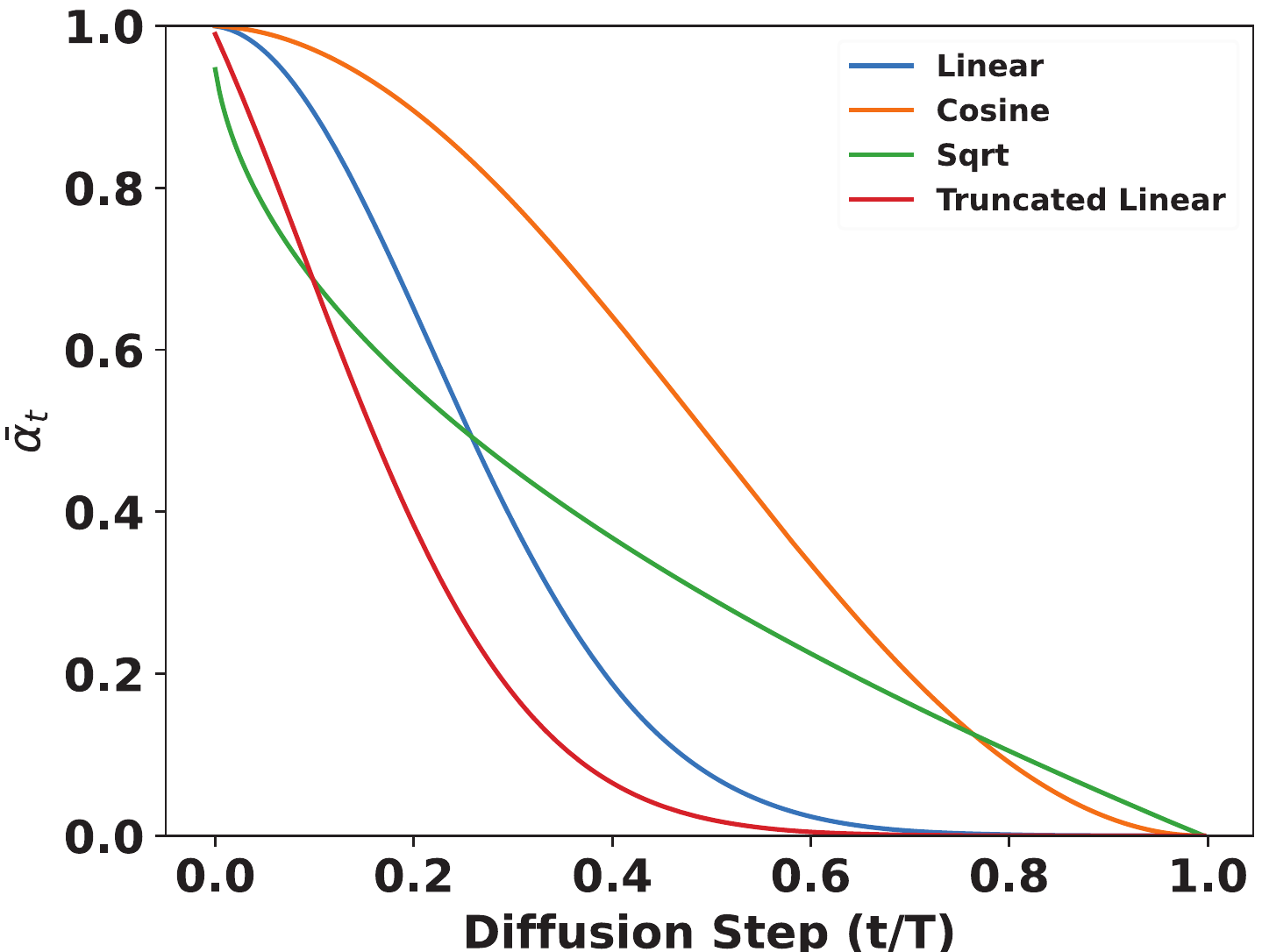}}
\caption{\textcolor{black}{$\bar{\alpha}_t$ throughout diffusion in different schedule.}}
\label{fig:alpha_t}
\end{figure}

\textcolor{black}{
Apart from that, we investigate the impact of different schedules to the final results. Following~\cite{nichol2021improved}, we plot the value of $\bar{\alpha}_t$ under different schedules as shown in Figure~\ref{fig:alpha_t}, where $\bar{\alpha}_t=\prod_{s=1}^t\alpha_s, \alpha_s=1-\beta_s$ (ref. Equation~\ref{eq:alpha_t}). We could find that the Truncated Linear schedule provides a near-linear sharp drop in the middle of the process and subtle changes near the extremes of $t=T$ compared with other schedules. This design proves to be effective in enhancing the model's capability of interest modeling since we aim to recover the original input from noise via only one-step estimation (ref. Algorithm~\ref{alg:reverse}). Consequently, a sharp corruption (\ie add more noise) in the training process is necessary for better denoising ability. However, we also find that in some cases the Linear schedule shows superiority over others, indicating some better alternatives that fall between Truncated Linear and Linear to be explored in the future.
Besides, as~\cite{he2022diffusionbert} emphasizes that different schedules will impact the final outcomes to some extent but could not acquire huge margin fluctuations. Our experiments also manifest this conclusion (see Figure~\ref{fig:schedule}) that the truncated linear schedule achieves the best results against other schedules on \textit{Amazon Beauty} and \textit{Amazon Baby} datasets but attain sub-optimal performance on the other two datasets.
} 
Moreover, following~\cite{li2022diffusion, ho2020denoisingdiff}, we explore the root mean-squared error loss (RMSE) and $\mathbf{\epsilon}$ prediction via \textit{Approximator} for target item representation reconstruction (\ie instead of predicted the target item representation $\mathbf{\hat{x}}$ straightforwardly. Specifically, we count on predicted $\mathbf{\epsilon}$ for target item representation reconstruction, $\mathbf{\hat{x}}_0=\frac{1}{\sqrt{\alpha_t}}(\mathbf{x_t}-\frac{\beta_t}{\sqrt{1-\overline{\alpha}_t}}\mathbf{\epsilon})$), but the training loss can not converge. Therefore, we do not report the final results here.

\begin{table}
\caption{\textcolor{black}{Results (\%) of ablation experiments. The best results are in boldface, and the second-best underlined.}}
\footnotesize
\begin{tabular}{clcccccc}
\hline
Dataset                       & Ablation    & HR@5    & HR@10   & HR@20   & NDCG@5 & NDCG@10 & NDCG@20 \\ \hline
\multirow{3}{*}{\rotatebox[origin=c]{90}{Beauty}}      
                              & w GRU      & \underline{3.1773}  & \underline{4.6685} & \underline{6.9000}  & \underline{2.2253} & \underline{2.7075}  & \underline{3.2682}  \\
                              & w R* & 1.3036  & 2.2902  & 3.4768 & 0.7964 & 1.1130  & 1.4108  \\
                              & \baby        & \textbf{5.5758}  & \textbf{7.9068}  & \textbf{11.1098}  & \textbf{4.0047} & \textbf{4.7494}  & \textbf{5.5566}  \\ \hline
\multirow{3}{*}{\rotatebox[origin=c]{90}{Toys}}      
                              & w GRU      & \underline{2.3895}  & \underline{3.4150} & \underline{4.9407}  & \underline{1.7577} & \underline{2.0849}  & \underline{2.4700}  \\
                              & w R* & 0.1194  & 0.4133  & 0.7480 & 0.0567 & 0.1510  & 0.2343  \\
                              & \baby        & \textbf{5.5650}  & \textbf{7.4587}  & \textbf{9.8417}  & \textbf{4.1667} & \textbf{4.7724}  & \textbf{5.3684}  \\ \hline     
\multirow{3}{*}{\rotatebox[origin=c]{90}{Movie}} 
                              & w GRU      & \underline{16.6016} & \underline{24.6094} & \underline{34.9772} &  \underline{10.9553} & \underline{13.5348}  & \underline{16.1403} \\
                              & w R* & 4.0690 & 6.0872 & 9.1688 & 2.7787 & 3.4297  & 4.1978 \\
                              & \baby        & \textbf{17.9659}  &  \textbf{26.2647}  & \textbf{36.7870} & \textbf{12.1150}  & \textbf{14.7909} & \textbf{17.4386} \\ \hline
\multirow{3}{*}{\rotatebox[origin=c]{90}{Steam}}      
                              & w GRU      & \underline{6.2832}  & \underline{10.1723} &  \underline{15.8169} & \underline{4.0056} & \underline{5.2548}  & \underline{6.6727}  \\
                              & w R* & 1.4873  & 2.5977  & 4.3086 & 0.9022 & 1.2582  & 1.6877  \\
                              & \baby        & \textbf{6.6742}  & \textbf{10.7520}  & \textbf{16.6507}  & \textbf{4.2902} & \textbf{5.5981}  & \textbf{7.0810}  \\ \hline
\end{tabular}
\label{tab:ablation}
\end{table}

\begin{figure*}[t]
\centerline{\includegraphics[width=0.95\textwidth]{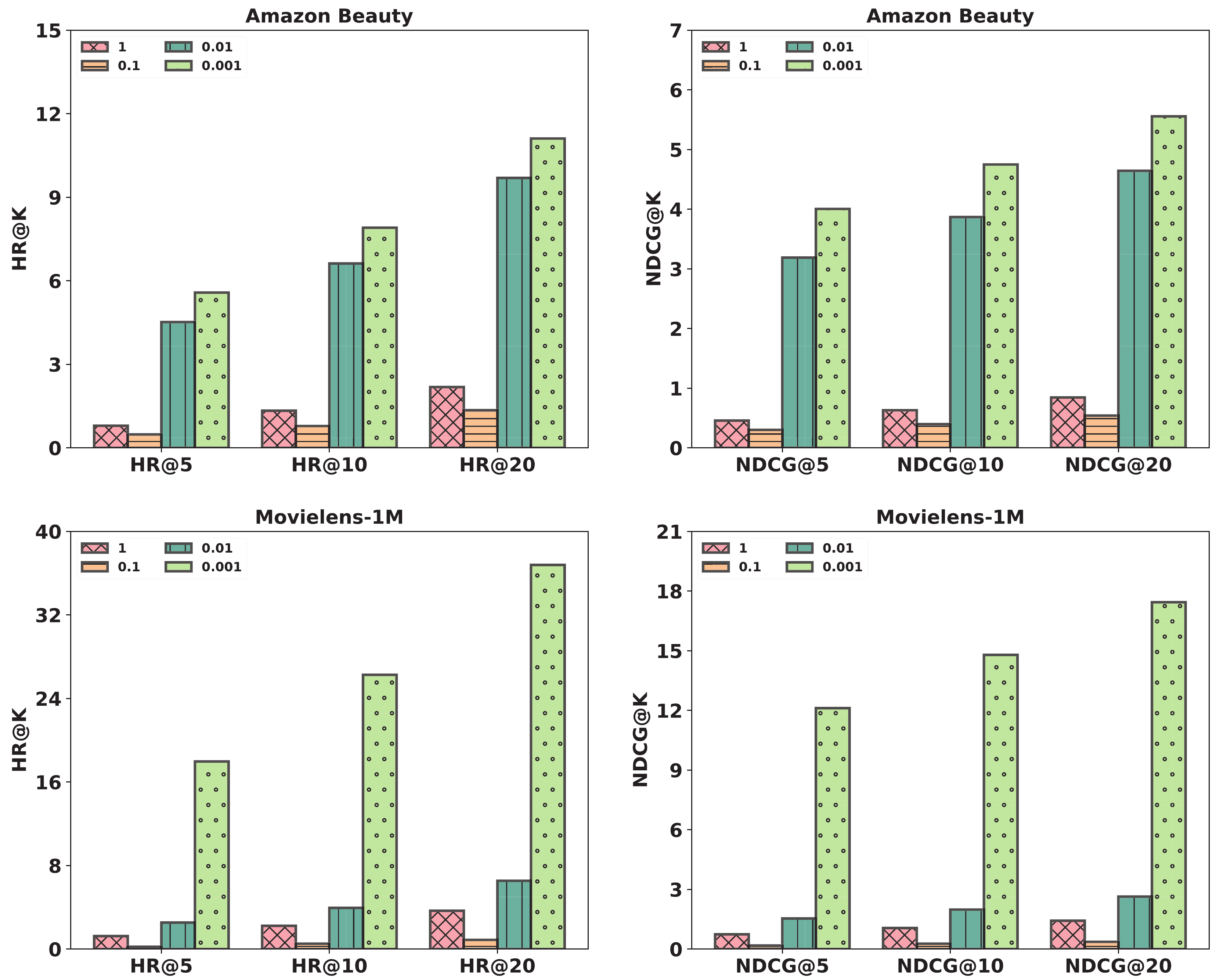}}
\caption{Parameter sensitivity of $\lambda$.}
\label{fig:hyperparameter}
\end{figure*}

\subsection{Impact of Hyper-parameter Setting (RQ3)}
\label{hyperparameter}
Besides the noise schedule, the value of $\lambda$ in Equation~\ref{eq:approximator} also regulates the discrimination capacity of the item representations. Here, we set different $\delta$ which controls the mean and variance of $\lambda$ sampling distribution and investigate the effect of this hyper-parameter on the performance of \baby. Figure~\ref{fig:hyperparameter} presents the performance patterns by varying $\lambda$ values. We can see that a small $\lambda$ generally resulted in better performance on both \textit{Amazon Beauty} and \textit{Movielens-1M} datasets, while when we increase the $\lambda$ to $0.1$, the \textit{HR} and \textit{NDCG} drop sharply, especially on \textit{Movielens-1M} dataset. We suppose that a large $\lambda$ may add too much noise to the historical interaction sequences, which will corrupt the original information and hinder the model to precisely understanding the user's interest.    

\subsection{Compared with Adversarial Training (RQ4)}
\label{sec:adv}

\begin{figure}[t]
\centerline{\includegraphics[width=\textwidth]{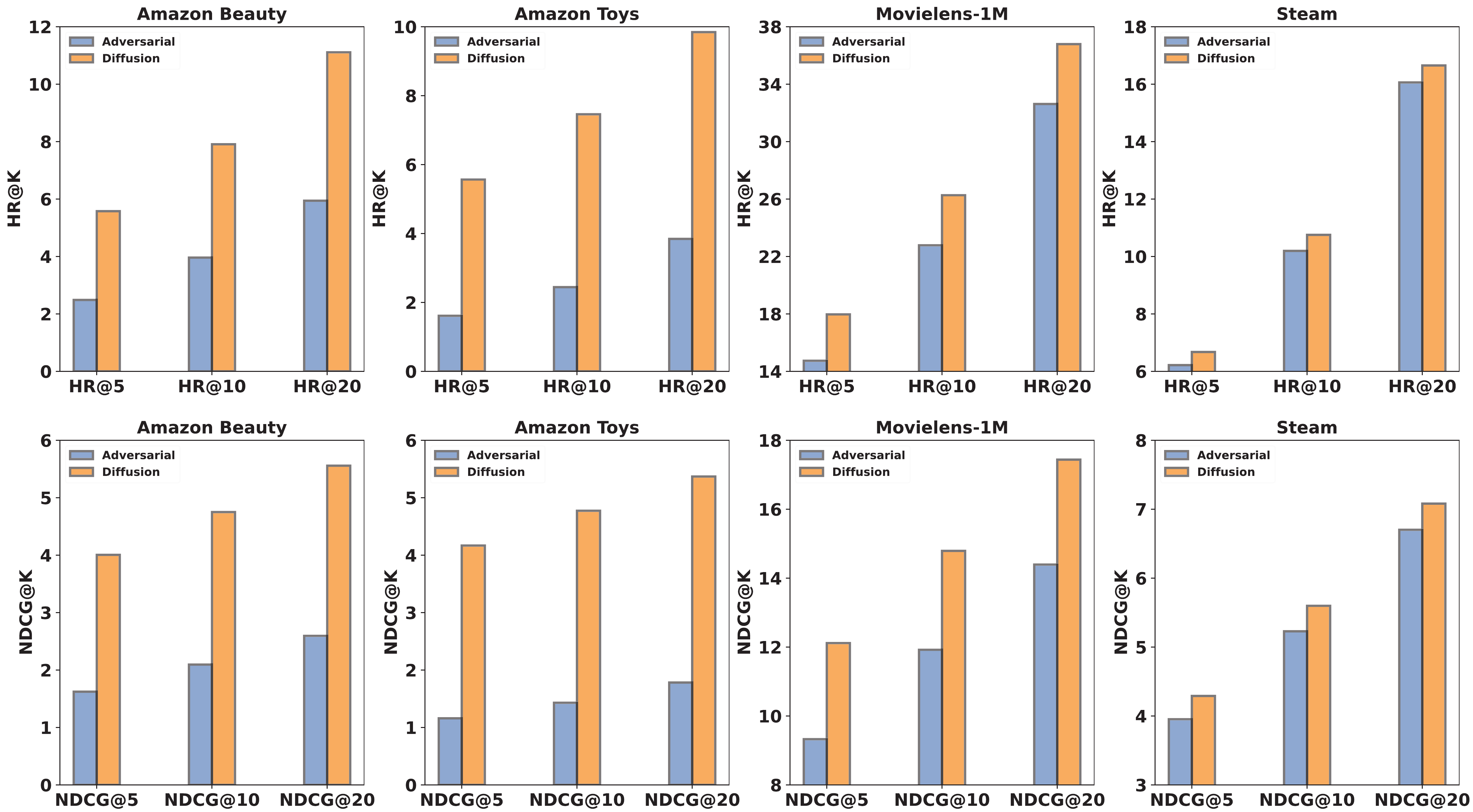}}
\caption{Performance of adversarial training and \baby.}
\label{fig:adv}
\end{figure}

To further analyze the effectiveness of uncertainty and diversity of diffusion model to the recommendation, we also adopt adversarial training for performance comparison. As described in Section~\ref{sec:discussion}, by adding adversarial perturbations on the training process, the model's robustness and performance can be improved. We 
set $\epsilon=0.5$ and $\gamma=1$. The results are shown in Figure~\ref{fig:adv}.

Observing Figure~\ref{fig:adv} we could find that the adversarial Transformer achieves better results against BERT4Rec, albeit the model architectures are the same. Specifically, adversarial Transformer improves the baseline on \textit{Amazon Beauty} and \textit{Movielens-1M} datasets w.r.t. \textit{HR} and \textit{NDCG} by at least $2.57\%/10.23\%$ and $8.05\%/4.89\%$, respectively. It proves that the perturbation on the adversarial training process will be helpful for model robustness and performance improvement. However, compared with \baby, our method is superior to the adversarial Transformer by a large margin on all the datasets. One explanation is that the injected uncertainty and perturbation in diffusion and reverse phases is not just a pure noise randomly sampled from a Gaussian distribution for model robustness argumentation, but some auxiliary information from target item representation will also be introduced as supervised signals to facilitate model to capture user's current intentions for final prediction, which may be the recipe of \baby could achieve best results. 


\subsection{Performance on Varying Lengths of Sequences and Items with Different Popularity (RQ5)}
\label{sec:longshort}

\begin{figure}[t]
\centerline{\includegraphics[width=\textwidth]{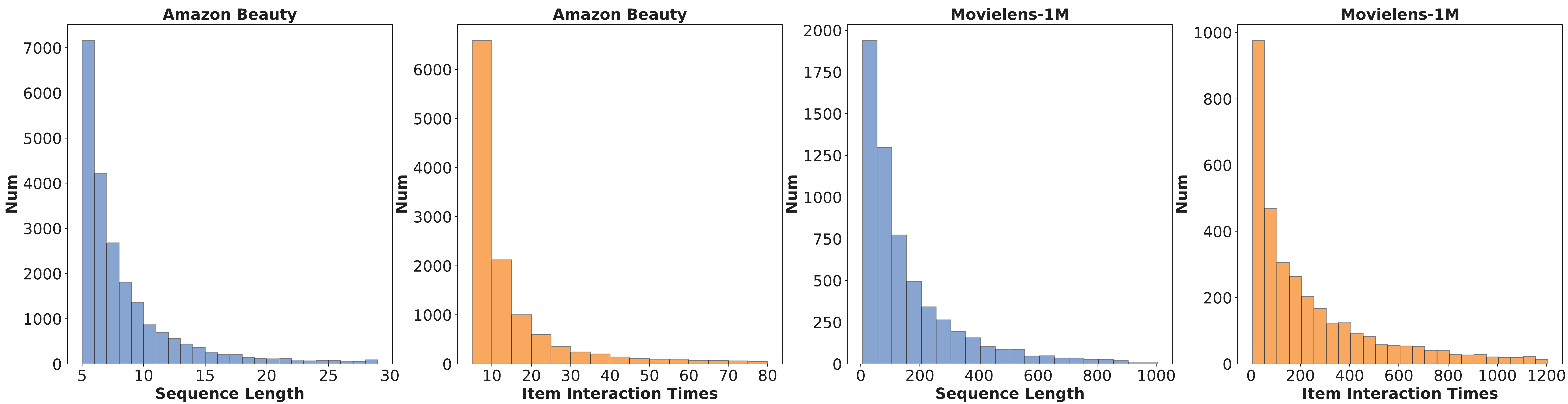}}
\caption{Frequency histogram of sequence length and item interaction times for \textit{Amazon Beauty} and \textit{Movielens-1M} dataset. The maximum length and maximum interactions are $204/431$ and $2341/3428$ respectively for \textit{Amazon Beauty} and \textit{Movielens-1M} datasets.}
\label{fig:long_head}
\end{figure}

We further analyze the performance of \baby against other competitive baselines (SASRec, TimiRec and STOSA) on different lengths of sequences as well as on head and long-tail items. Figure~\ref{fig:long_head} shows the distribution of sequence length and item interaction times on \textit{Amazon Beauty} and \textit{Movielens-1M} datasets, respectively. We could find that for \textit{Amazon Beauty} dataset, the length is less than $15$ for most of the sequences and the interaction times are less than $40$ for most of the items. So as to the \textit{Movielens-1M} dataset, although the sequence length and interaction numbers far exceed \textit{Amazon Beauty} dataset, the short sequence and long-tail items are still the majority. 

\begin{figure}[t]
\centerline{\includegraphics[width=\textwidth]{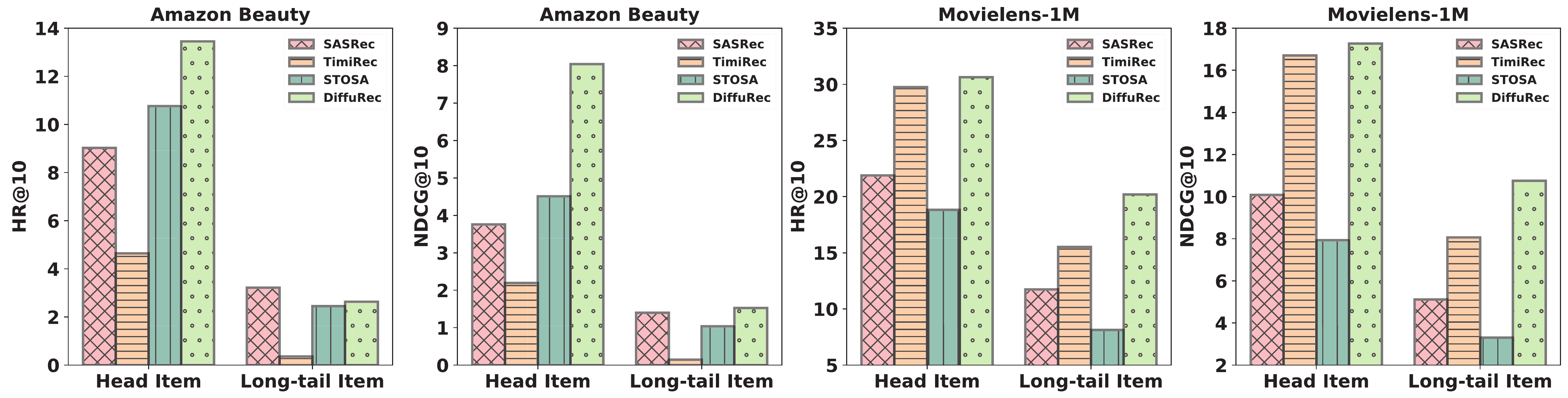}}
\caption{Performance on the head and long-tail items.}
\label{fig:head_tail}
\end{figure}

\paratitle{Head and Long-tail Items.} Following previous works~\cite{zhang2021model}, we split the first $20\%$ most frequent items as head items and denote the rest are long-tail items, guided by Pareto Principle~\cite{box1986analysis}, for performance evaluation. The results are shown in Figure~\ref{fig:head_tail}. We could find that for both \textit{Amazon Beauty} and \textit{Movielens-1M} datasets, the performance on head items is significantly superior to the long-tail items for all the models. 
However, this gap is notably smaller on the \textit{Movielens-1M} dataset, where the performance difference between head items and long-tail items is less than three times that of \textit{Amazon Beauty} dataset, where it is more than five times greater. Given that the number of items on \textit{Movielens-1M} is rather small, this might be the reason why there is no obvious performance difference between head items and long-tail items. Overall, \baby outperforms all baselines on most of the settings, in particular in \textit{NDCG} metric.

\begin{figure}[t]
\centerline{\includegraphics[width=\textwidth]{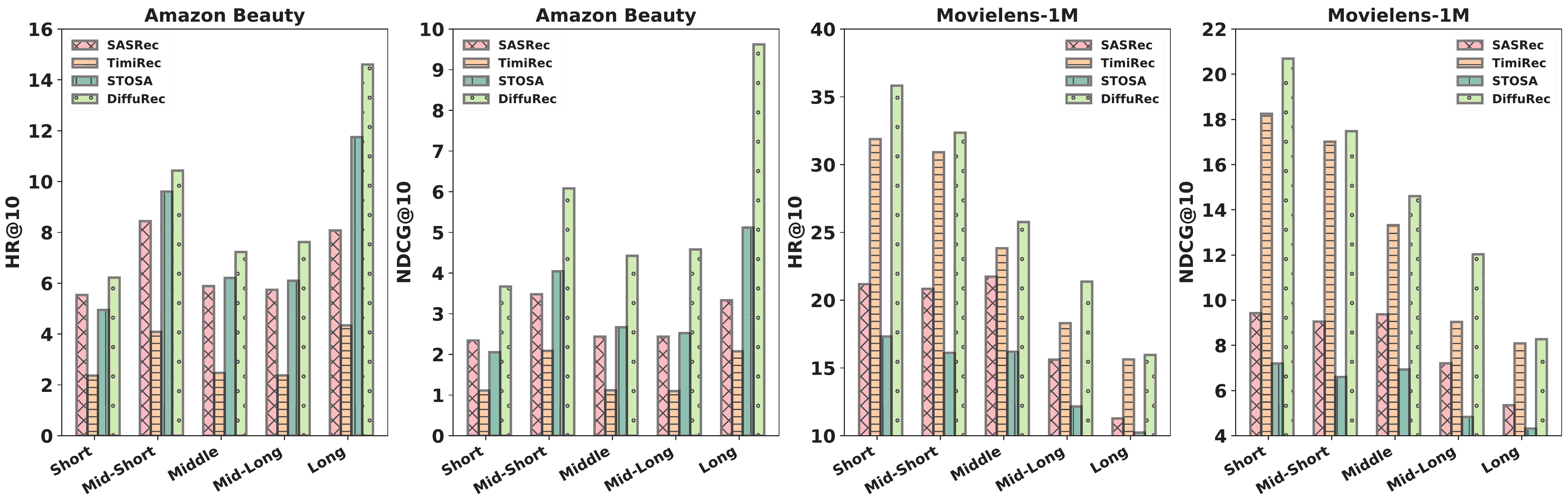}}
\caption{Performance on different length of sequence.}
\label{fig:length}
\end{figure}

\paratitle{Different Sequence Length.} Based on the percentile of sequence length $l$, we split the \textit{Amazon Beauty} / \textit{Movielens-1M} datasets into five groups: short ($0<l\leq 5$ / $0<l\leq37$), mid-short ($5<l\leq 6$ / $37<l\leq 70$), medium ($6<l\leq7$ / $70<l\leq 126$), mid-long ($7<l\leq 10$ / $126<l\leq 253$), and long ($l > 10$ / $l>253$). The performance of different baselines on varied sequence lengths is depicted in Figure~\ref{fig:length}, it can be observed that the results on \textit{Movielens-1M} dataset and \textit{Amazon Beauty} dataset are completely opposite. As for \textit{Amazon Beauty} dataset, \baby achieves a better performance on long sequences. Conversely, for the \textit{Movielens-1M} dataset, as the sequence length increases, there will be a gradual decline in performance. Comparing the sequence length of these two datasets, we could find that $80\%$ of sequences are shorter than $10$ for \textit{Amazon Beauty} dataset whereas for the \textit{Movielens-1M} dataset $80\%$ of sequences are longer than $37$. 
We believe the short sequences can not provide sufficient information to reveal the user's current preference for a precise recommendation, whereas it is also challenging for all the models to handel very long sequences and recommendation.  In general, compared with different baselines, our model achieves the best results in all the settings, which manifests that \baby is robustness to the sequence length. 

\subsection{Convergence and Efficiency (RQ6)}
\label{efficiency}

\begin{figure}[t]
\centerline{\includegraphics[width=0.95\textwidth]{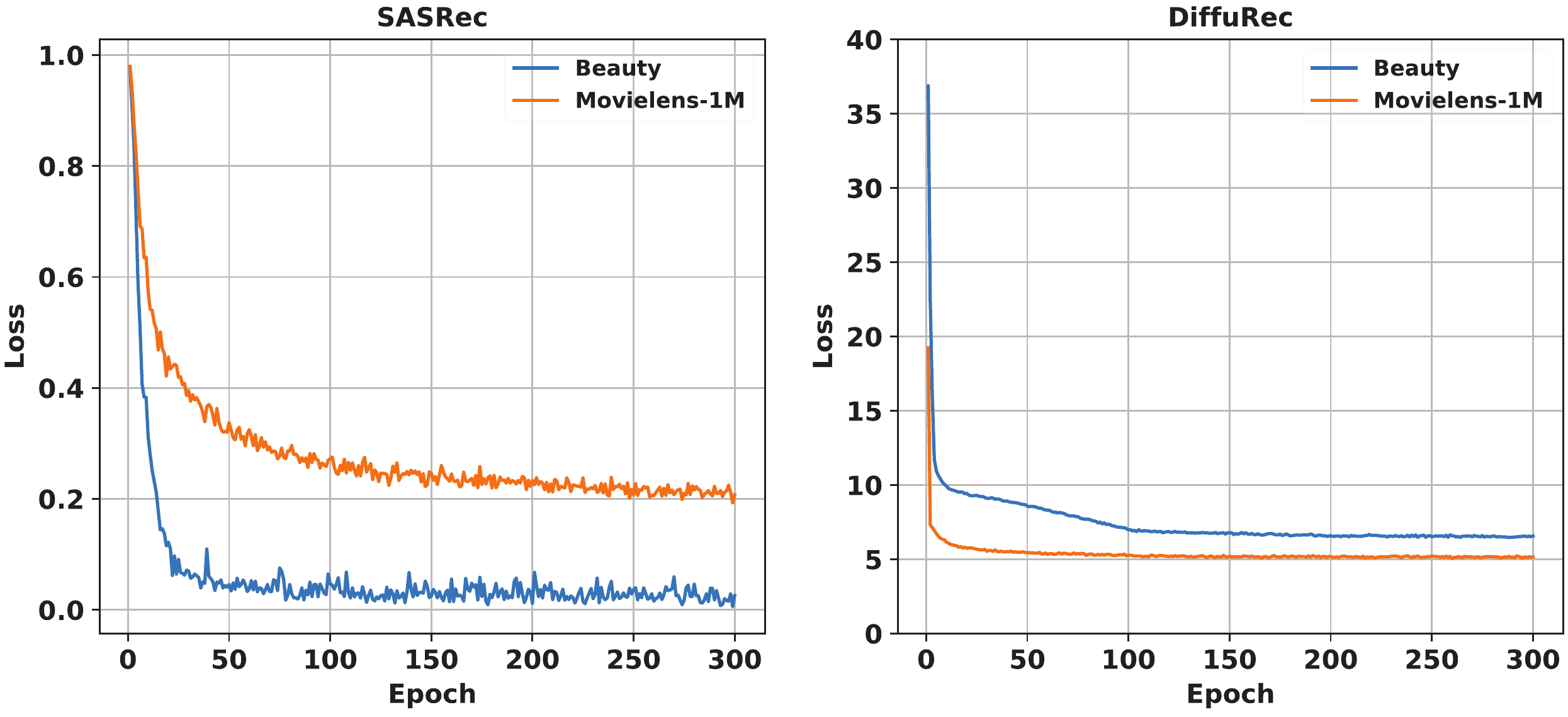}}
\caption{Curve of training loss on \textit{Amazon Beauty} and \textit{Movielens-1M} datasets}
\label{fig:convergence}
\end{figure}

\baby is a new paradigm for sequential recommendation which is also an iterative process in reverse and model inference. We thereby analyze its convergence on training process and efficiency for inference. All experiments are conducted on an NVIDIA GeForce RTX 3090 GPU and Intel Xeon CPU E5-2680 v3.

\paratitle{Convergence.} We compare the convergence speed of our model with SASRec, a representative sequential recommendation baseline, whose model structures are also similar to ours (\ie Transformer-based). For a fair comparison, we keep the optimizer, dropout ratio, hidden size, multi-head numbers, and Transformer block numbers as the same, \ie Adam, $0.1, 128, 4$, and $4$ respectively.  

Figure~\ref{fig:convergence} depicts the loss curves on training process. Observe that on \textit{Amazon Beauty} dataset both \baby and SASRec could converge after $150$ training epochs. But on \textit{Movielens-1M} dataset, \baby presents faster convergence (around 100 epochs) than SASRec (around $250$ epochs). We speculate it is because the average length of \textit{Movielens-1M} dataset ($95.19$) is over ten times than \textit{Amazon Beuaty} dataset ($8.53$), SASRec thereby could not capture the long-term dependency between items efficiently, while our model can induce the user's current interest extracted from the target item representation as auxiliary information, which accelerates the training convergence speed of models.

\begin{figure}
\centering
\begin{minipage}[t]{0.65\textwidth}
\centering
\includegraphics[height=2.4in]{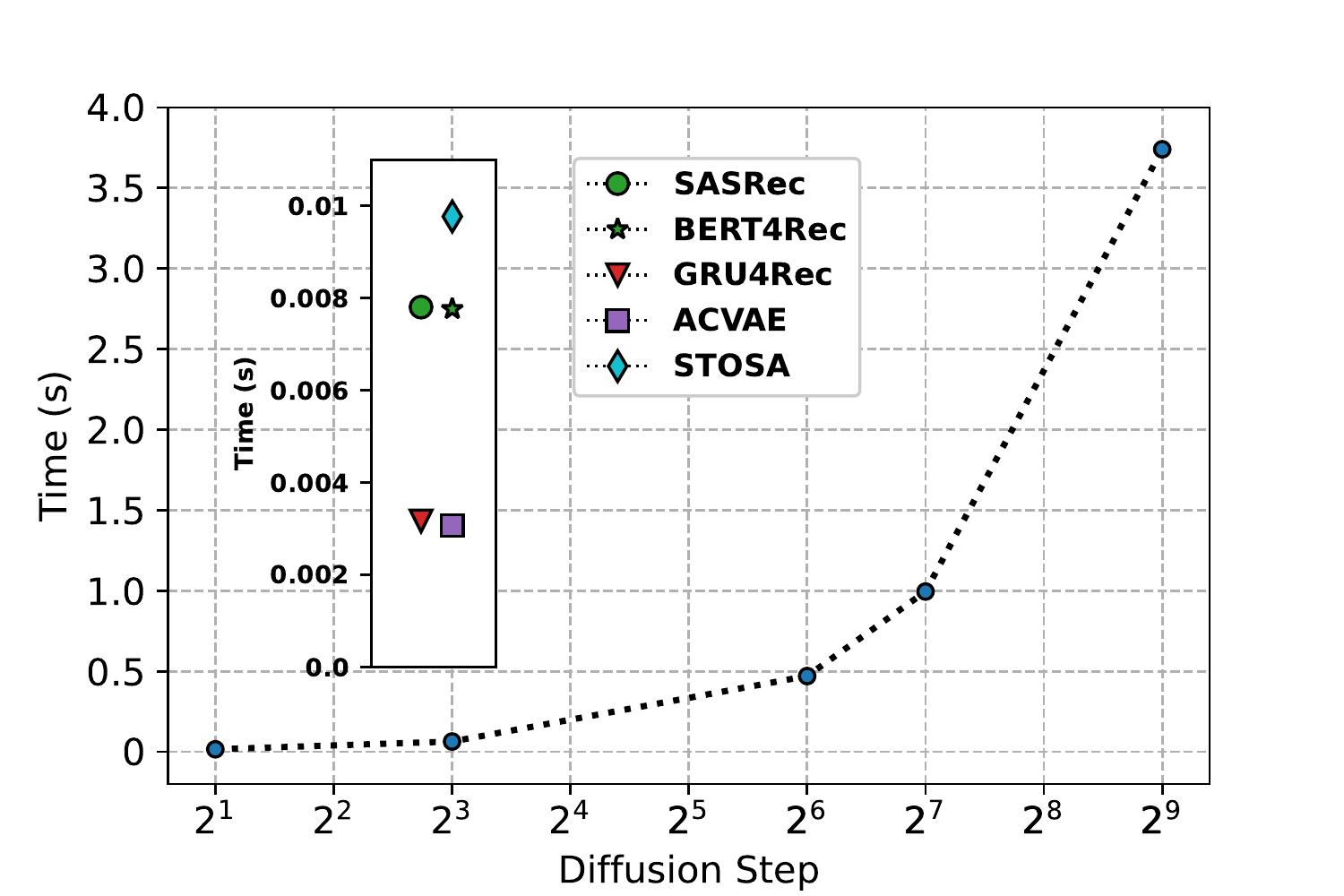}
\end{minipage}
\begin{minipage}[t]{0.32\textwidth}
\centering
\includegraphics[height=2.2in]{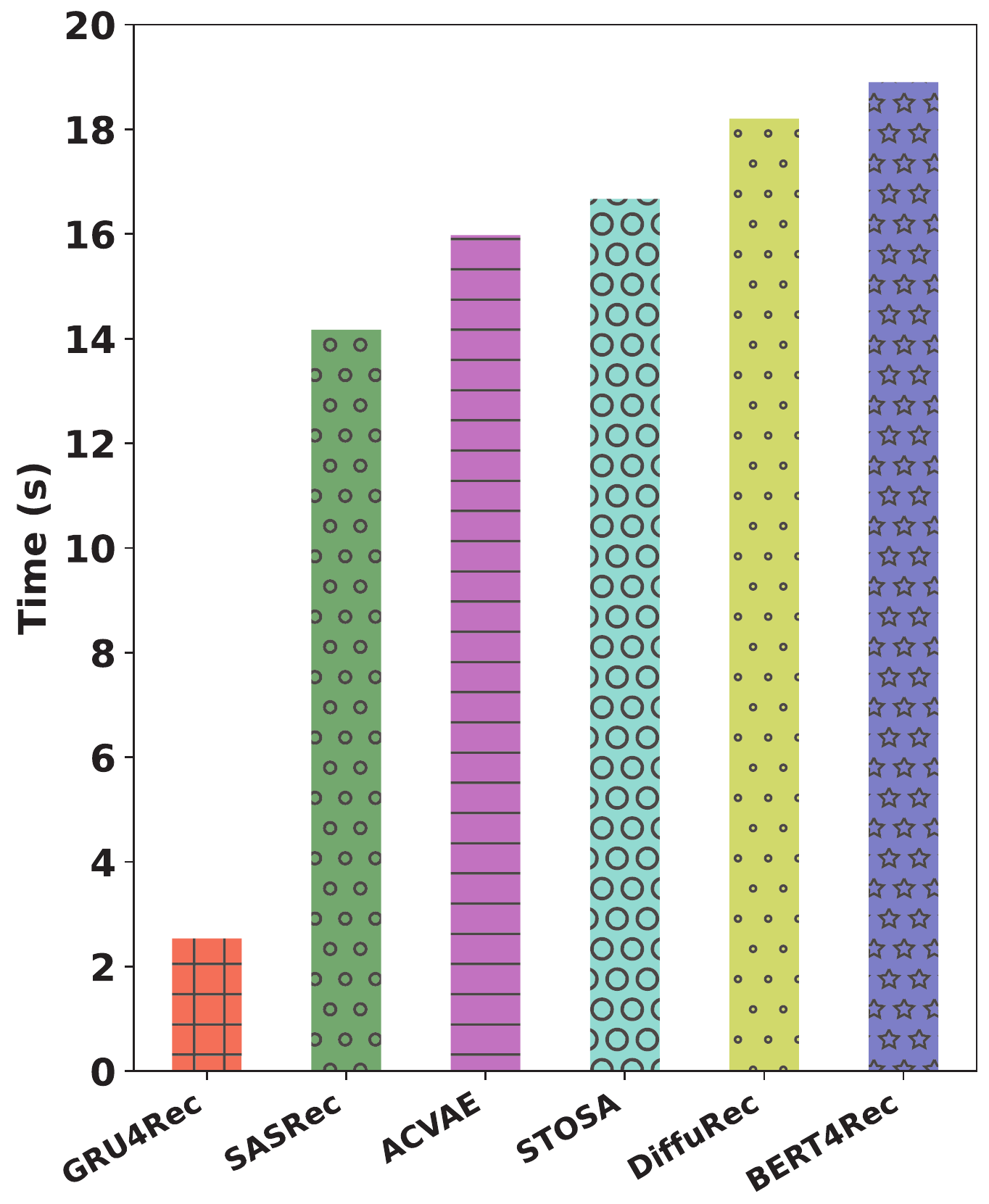}
\end{minipage}
\caption{\textcolor{black}{Inference time (in seconds for one sample) on \textit{Amazon Beauty} dataset (left). The points in the inset box present the inference time of baselines. The folding line chart reveals the inference time for the different reverse steps. Average training time (in seconds for one batch) on \textit{Amazon Beauty} dataset (right).}}
\label{fig:effiency}
\end{figure}

\paratitle{Efficiency.} In general, the number of reverse steps is the most critical factor to determine the inference speed of diffusion models. \textcolor{black}{Hence, we investigate the effect of reverse step to inference time and also analyze the time complexity with big O notation of \baby against other recipes. The results are shown in Figure~\ref{fig:effiency} and Table~\ref{tab:complexity}.}

\begin{table}
\caption{\textcolor{black}{Time complexity of \baby and other baselines. $d$ and $n$ are the representation dimension and sequence length respectively.}}
\begin{tabular}{lcccccc}
\hline
Model      & GRU4Rec                  & SASRec                                         & BERT4Rec                                       & ACVAE                    & STOSA                                          & DiffuRec                                       \\ \hline
Complexity & $O(nd^2)$ & $O(nd^2+dn^2)$ &$O(nd^2+dn^2)$  & $O(nd^2)$ & $O(nd^2+dn^2)$  & $O(nd^2+dn^2)$  \\ \hline
\end{tabular}
\label{tab:complexity}
\end{table}

\begin{figure}[t]
\centerline{\includegraphics[width=\textwidth]{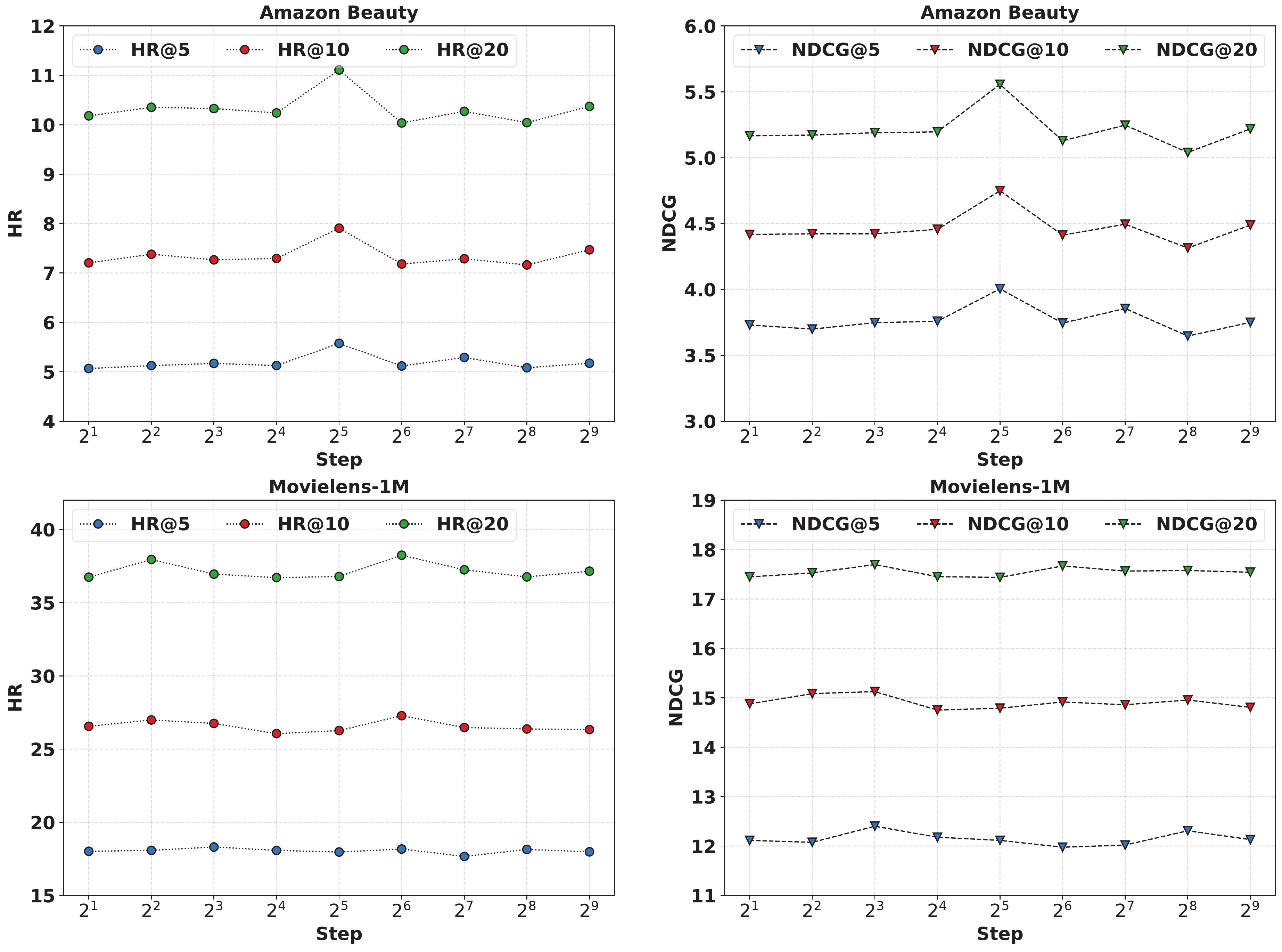}}
\caption{Recall@K and NDCG@K at different reverse steps on \textit{Amazon Beauty} and \textit{Movielens-1M} datasets.}
\label{fig:performance}
\end{figure}

Observing Figure~\ref{fig:effiency} (left), the inference time of \baby is close to SASRec and BERT4Rec when we set the total diffusion step as a small value (\ie $2$), as all of them are Transformer-based models and the time complexity is the same (ref. Tabel~\ref{tab:complexity}). Besides, with the increase of reverse step, the inference time for one sample rises exponentially, while the performance also exhibits a fluctuation (observing Figure~\ref{fig:performance}).
Balancing the quality of representation and the time, we consider a moderate reverse step (\eg 32 or 64) is sufficient for SR. Furthermore, compared with Transformer-based methods, both GRU4Rec and ACVAE apply GRU as the backbone for recommendation. Hence, their inference time and time complexity are similar and shorter than Transformer-based solutions. STOSA improves Transformer as a stochastic model and also adopts the Wassertein distance as an alternative to the inner product for discrepancy measure between two items. Consequently, the inference time is slightly longer than the unmodified Transformer-based methods.

\textcolor{black}{We further compare the average training time of one epoch  between \baby and other baselines, as shown in Figure~\ref{fig:effiency} (right). We could find that the training time of GRU4Rec is significantly less than others since the model structure and size are more lightweight than the Transformer-based architectures. Noting that the training time of SASRec, ACVAE, STOSA, \baby, and BERT4Rec are very close, with all models encompassing a range from $14s$ to $18s$. Thus, we believe the training process of our \baby will not be the bottleneck of model optimization.}

\subsection{\textcolor{black}{Visualization of Uncertainty and Diversity (RQ7)}} 
\label{efficiency}

\begin{figure}[t]
\centerline{\includegraphics[width=\textwidth]
{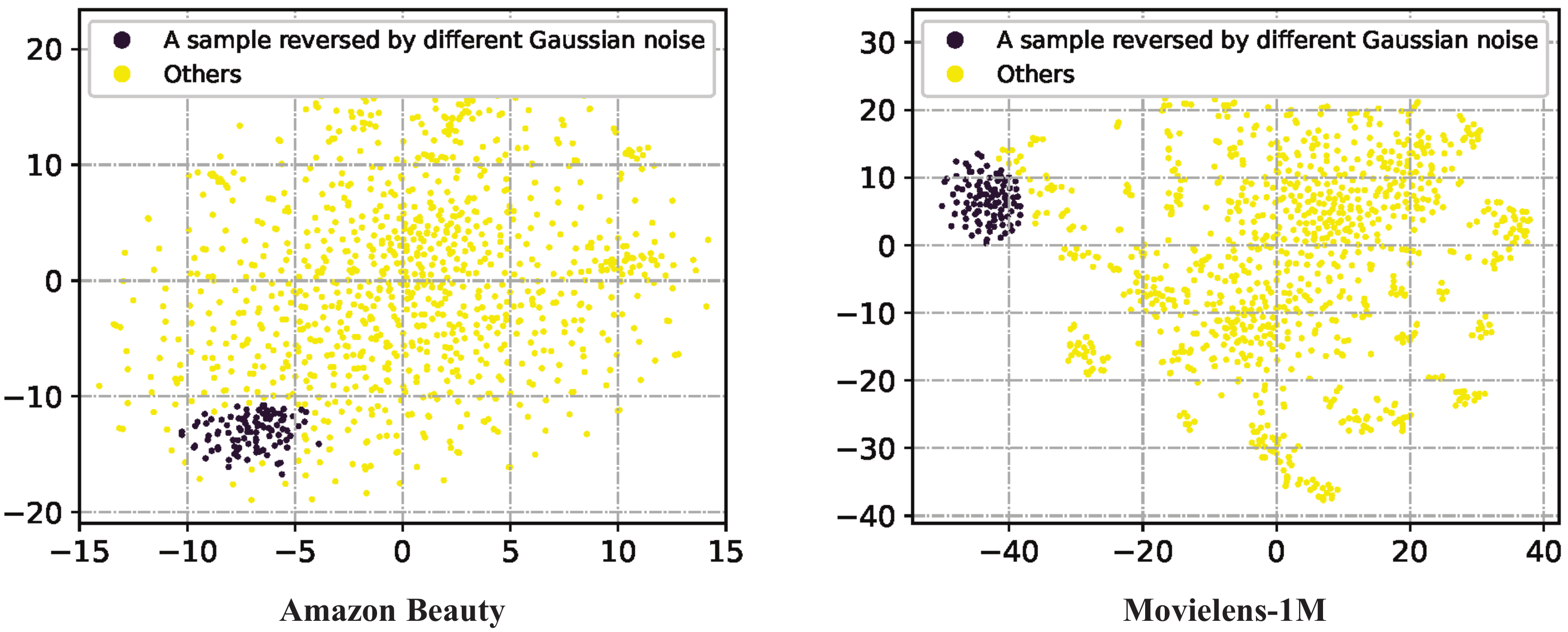}}
\caption{A t-SNE plot of the reconstructed $\mathbf{x}_0$ representation from reverse phase. The purple points represent the same item but reversed by $100$ different Gaussian noise for a specific historical sequence. The yellow points are the others each of which is reversed from a random sequence in \textit{Amazon Beauty} and \textit{Movilens-1M} datasets respectively.}
\label{fig:uncertainty}
\end{figure}

\paratitle{\textcolor{black}{Uncertainty.}} As we emphasize that attributed to the theoretical underpinnings of Diffusion framework, \baby could inject some uncertainty in diffusion and reverse process, which will be beneficial to the final performance improvement. We then analyze the uncertainty of the reconstructed target item representation $\mathbf{x}_0$ reversed by our method. Figure~\ref{fig:uncertainty} plots a set of reversed $\mathbf{x}_0$ vectors and other item representation from \textit{Amazon Beauty} and \textit{Movielens-1M} datasets projected into a two-dimensional space via t-SNE~\cite{van2008visualizing}.

Observing Figure~\ref{fig:uncertainty}, we find that the reversed representation $\mathbf{x}_0$ from random sequences are dispersed throughout the whole space uniformly. 
In contrast, the reversed $\mathbf{x}_0$ of the exemplar sequence under different Gaussian noises are relatively much closer but also hold certain deviations to each other. Actually, most existing solutions are deterministic for a specific sequence.
The merit of this uncertainty is much desired for the retrieval stage. Specifically, by reversing $100$ different $\mathbf{x}_0$ for the exemplar sequence, the number of unique items that are ranked in their top-$20$ reaches $643$ and $82$ respectively for \textit{Amazon Beauty} and \textit{Movielens-1M} datasets.
That is, we can generate different reversed $\mathbf{x}_0$ with parallel computing, consequently leading to the diversified retrieval results.   

\paratitle{\textcolor{black}{Diversity.}}\textcolor{black}{To verify the impact of uncertainty on the diversity of final recommendation results, we randomly sample one sequence from \textit{Amazon Toys} dataset and collect the categories of the recommended items. Table~\ref{tab:diversity} depicts the top five items' categories yielded by our \baby and other alternatives.}


\begin{table}
\caption{\textcolor{black}{The category of recommended Top 5 items. Items sharing the same color indicate they are in similar categories.}}
\footnotesize
\setlength\tabcolsep{1.5pt}
\begin{tabular}{llc}
\hline
\textbf{Model} & \multicolumn{1}{c}{\textbf{Categories of Top 5 Items.}}                                                                                                                                         & \textbf{Target Item Category} \\ \hline
SASRec         & {\begin{tabular}[c]{@{}l@{}}\textcolor{black}{Kids' Electronics}, \textcolor{black}{Electronic Toys}, \textcolor{orange}{Learning \& Education}, \\ \textcolor{black}{Toy Remote Control \& Play Vehicles}, \textcolor{black}{Play Vehicles}\end{tabular}}     & Electronic Toys        \\\hline
TimiRec        & \begin{tabular}[c]{@{}l@{}}\textcolor{black}{Toy Remote Control \& Play Vehicles}, \textcolor{black}{Electronic Toys}, \textcolor{black}{Play Vehicles}, \\ \textcolor{black}{Remote \& App Controlled Vehicle Parts}, \textcolor{black}{Play Trains \& Railway Sets}\end{tabular} & Electronic Toys        \\\hline
STOSA          & \textcolor{orange}{Learning \& Education}, \textcolor{black}{Playsets \& Vehicles}, \textcolor{purple}{Board Games}, \textcolor{black}{Train Sets}, \textcolor{brown}{Wind-up Toys}   & Electronic Toys       \\\hline
\baby      & \begin{tabular}[c]{@{}l@{}}\textcolor{black}{Electronic Toys}, \textcolor{black}{Helicopter Kits}, \textcolor{black}{Remote \& App Controlled Vehicles}, \\ \textcolor{orange}{Card Games}, \textcolor{purple}{Electronic Software \& Books}\end{tabular}                                          & Electronic Toys        \\ \hline
\end{tabular}
\label{tab:diversity}
\end{table}

\textcolor{black}{As shown in Table~\ref{tab:diversity}, SASRec and TimiRec are inclined to recommend items from similar categories, \eg Electronic Toys, and Vehicles, given that TimiRec leverages the target item as the supervised signal for recommendation, which will constrain the final outcomes to the similar categories inherently. On the contrary, attributed to the uncertainty injection, STOSA, and \baby could recommend a wider range of categories rather than focusing on the "Electronic Toys". As such, we believe injecting some uncertainty in recommendation will facilitate more flexibility and therefore yield more diversity and richer recommendation results.}

\section{Conclusion}
\label{sec:conclusion}

This work focuses on sequential recommendation. We consider that the single vector based representation used in current solutions does not well capture the dynamics in the recommendation context. To model items' latent aspects and users' multi-level interests, we consider Diffusion model to be a good fit. 
To instantiate this idea, we adapt the Diffusion model to sequential recommendation for the very first time and propose \baby, which is our key contribution. Specifically, we detail how the diffusion and reverse phase are carried out under the new problem setting, and how to design an Approximator. 
Experimental results demonstrate the superiority of \baby on four real-world datasets.
We have also conducted extensive experiments to verify the effectiveness of the designed components in \baby. 
As a new paradigm to recommendation tasks, different ways \textcolor{black}{and other recommendation scenarios (\eg session-based recommendation, click-through rate prediction)} of adapting the powerful Diffusion model remain under-explored. 
We hope this work may shed light along this direction.

\bibliographystyle{ACM-Reference-Format}
\bibliography{bib}

\end{document}